\documentclass[]{aastex631}

\usepackage[T1]{fontenc}
\usepackage{soul}
\usepackage{mathtools}

\shorttitle{Why Enceladus and Europa's ice shell look so different?}
\shortauthors{Kang}
\graphicspath{{./}{figures/}}

\begin{document}

\title{Different ice shell geometries on Europa and Enceladus due to their different sizes: impacts of ocean heat transport.}

\correspondingauthor{Wanying Kang}
\email{wanying@mit.edu}

\author[0000-0002-4615-3702]{Wanying Kang}
\affiliation{Earth, Atmospheric and Planetary Science Department, 
  Massachusetts Institute of Technology,
  Cambridge, MA 02139, USA}




\begin{abstract}
On icy worlds, the ice shell and subsurface ocean form a coupled system -- heat and salinity flux from the ice shell induced by the ice thickness gradient drives circulation in the ocean, and in turn, the heat transport by ocean circulation shapes the ice shell. Therefore, understanding the dependence of the efficiency of ocean heat transport (OHT) on orbital parameters may allow us to predict the ice shell geometry before direct observation is possible, providing useful information for mission design. Inspired by previous works on baroclinic eddies, I first derive scaling laws for the OHT on icy moons, driven by ice topography, and then verify them against high resolution 3D numerical simulations. Using the scaling laws, I am then able to make predictions for the equilibrium ice thickness variation knowing that the ice shell should be close to heat balance. Ice shell on small icy moons (e.g., Enceladus) may develop strong thickness variations between the equator and pole driven by the polar-amplified tidal dissipation in the ice, to the contrary, ice shell on large icy moons (e.g., Europa, Ganymede, Callisto etc.) tends to be flat due to the smoothing effects of the efficient OHT. These predictions are manifested by the different ice evolution pathways simulated for Enceladus and Europa, considering the ice freezing/melting induced by ice dissipation, conductive heat loss and OHT as well as the mass redistribution by ice flow.
\end{abstract}



\section{Introduction}
Many of the icy satellites in the outer solar system are likely to contain a subsurface ocean underneath their ice shell due to tidal dissipation \citep{Scharf-2006:potential}, which may lead to a suitable environment for life to thrive. Enceladus (a satellite of Saturn) and Europa (a satellite of Jupiter), in particular, have been confirmed to have a global subsurface ocean by data brought back by the Galileo and Cassini missions \citep{Postberg-Kempf-Schmidt-et-al-2009:sodium, Thomas-Tajeddine-Tiscareno-et-al-2016:enceladus, Carr-Belton-Chapman-et-al-1998:evidence, Kivelson-Khurana-Russell-et-al-2000:galileo, Hand-Chyba-2007:empirical}. 
As two of the most enigmatic targets to search for extraterrestrial life \citep{Des-Nuth-Allamandola-et-al-2008:nasa, Hendrix-Hurford-Barge-et-al-2019:nasa}, Enceladus and Europa are to be further explored in the future (e.g., \textit{Europa Clipper} and \textit{JUICE}). Thus far, the measurements and detection have be carried out above the ice shell for the most part, and are likely to remain that way in the near future. Therefore, it becomes crucial to infer the subsurface condition using the information collected above the surface and to put better constraints on the ice shell geometry so we know where to send our landing or drill missions.

These motivate us to study the interaction between the subsurface ocean and the ice shell that we can more easily take measurement of. As illustrated by \citet{Kang-Mittal-Bire-et-al-2021:how} and \citet{Kang-Jansen-2022:in}, the interaction happens in a mutual way: the variation of ice thickness on Enceladus can drive ocean circulation by inducing salinity flux through freezing/melting and by changing the local freezing point; in turn, ocean circulation can converge heat to regions covered by relatively thick ice, flattening the ice shell (sketched in Fig.~\ref{fig:EOS-Hice-Heatflux}d). Such interaction makes it possible to infer the subsurface ocean properties from the information about the ice shell geometry or to infer the ice shell geometry based on our understanding of ocean dynamics \citep{Kang-Jansen-2022:in}.

Thus far, based on the surface measurements (libration, shape, gravity etc.) done by Cassini, Enceladus' ice shell has been revealed to be around 20~km thick on global average and to become significantly thinner toward the poles \citep{Iess-Stevenson-Parisi-et-al-2014:gravity, Beuthe-Rivoldini-Trinh-2016:enceladuss, Tajeddine-Soderlund-Thomas-et-al-2017:true, Cadek-Soucek-Behounkova-et-al-2019:long, Hemingway-Mittal-2019:enceladuss}. Whilst the equatorial ice shell is 30~km thick, the ice shell thickness over the south pole is only 6~km \citep{ Hemingway-Mittal-2019:enceladuss}; all geysers gather around this area, opening up periodically under tidal stress \citep{Hedman-Gosmeyer-Nicholson-et-al-2013:observed, Hurford-Helfenstein-Hoppa-et-al-2007:eruptions, Nimmo-Porco-Mitchell-2014:tidally, Ingersoll-Ewald-Trumbo-2020:time} and ejecting samples of the ocean to outer space \citep{Porco-Helfenstein-Thomas-et-al-2006:cassini, Hansen-Esposito-Stewart-et-al-2006:enceladus, Howett-Spencer-Pearl-et-al-2011:high, Spencer-Howett-Verbiscer-et-al-2013:enceladus}. In order to sustain the strong ice topography, the ocean heat transport (OHT) that flattens the ice shell through ice pump mechanism \citep{Lewis-Perkin-1986:ice} cannot be arbitrarily strong, which in turn puts constraints on the ocean salinity and the partition of heat production between the silicate core and the ice shell \citep{Kang-Mittal-Bire-et-al-2021:how}.

Europa's ice shell geometry is not as well constrained, but evidence has been found in favor a relatively thin \citep[$<$15~km][]{Hand-Chyba-2007:empirical} and flat \citep{Nimmo-Thomas-Pappalardo-et-al-2007:global} ice shell, in line with the evidence that Europa geysers are not as concentrated \citep{Roth-Saur-Retherford-et-al-2014:transient, Jia-Kivelson-Khurana-et-al-2018:evidence, Arnold-Liuzzo-Simon-2019:magnetic, Huybrighs-Roussos-Bloecker-et-al-2020:active}. As a separate line of evidence, \citet{Kang-Jansen-2022:in} found that the OHT-induced per-area heat flux scales with the satellite's radius to the power of 0.5 or 1 (depending on whether the magnitude of vertical diffusivity is sufficient to communicate the entire ocean column from the ice to the seafloor), which also supports a rather flat ice shell on Europa given its large size. 

However, what has been missing in the framework proposed by \citet{Kang-Mittal-Bire-et-al-2021:how} and \citet{Kang-Jansen-2022:in} are eddy dynamics. We know from earth ocean that ocean can be filled by eddies of various scales and that they play an important role in transporting heat, momentum and tracers \citep{Thompson-2008:atmospheric, Volkov-Lee-Fu-2008:eddy, Thompson-Heywood-Schmidtko-et-al-2014:eddy}. Scaling laws that govern the eddy diffusivity, tracer transports and equilibrium stratification have been found under forcings that are relevant for earth ocean or atmosphere \citep{Held-Larichev-1996:scaling, Karsten-Jones-Marshall-2002:role, Jansen-Ferrari-2013:equilibration}. In the context of icy satellites, recent works by \citet{Ashkenazy-Tziperman-2021:dynamic} and \citet{Ashkenazy-Tziperman-2016:variability} demonstrate that eddies also exist in the subsurface ocean of snowball Earth and Europa, and a set of sensitivity tests run under 3D configuration in \citet{Kang-Mittal-Bire-et-al-2021:how} also shows strong baroclinic eddies. Finding scaling laws for the eddy heat transport in icy moon oceans and making predictions about the equilibrium ice shell geometry are the main goals of this work.

The works on the eddy transports across the the antarctic circumpolar current (ACC) on earth provide useful insights for the OHT scaling on icy moons. However, a few differences between icy moon oceans and the terrestrial ocean should be noted. The overturning circulation near the ACC is largely driven by surface wind stress especially: curl of wind stress forces meridional overturning flows, maintaining isopycnal slope so that eddies can grow on it \citep{McWilliams-Holland-Chow-1978:description, Marshall-Radko-2003:residual, Plumb-Ferrari-2005:transformed}. To the contrary, the ocean on an icy satellite is sandwiched between an ice shell and a silicate core, and thus experience no wind stress. Density gradients created by the ocean-ice heat and salinity exchange, instead, drive the overturning circulation there. Since both the top and bottom boundaries are frictious on icy satellites, they can provide the drag required to balance the Coriolis acceleration associated with the overturning flows along the boundaries \citep{Kang-Jansen-2022:in}. The equilibrium temperature and salinity profiles adjusted only by the zonally-symmetric overturning circulation may undergo baroclinic instability \citep{Charney-1947:dynamics}, which further boosts the OHT.


\section{The coupled ocean-ice system.}

\begin{figure*}
    \centering \includegraphics[page=1,width=0.9\textwidth]{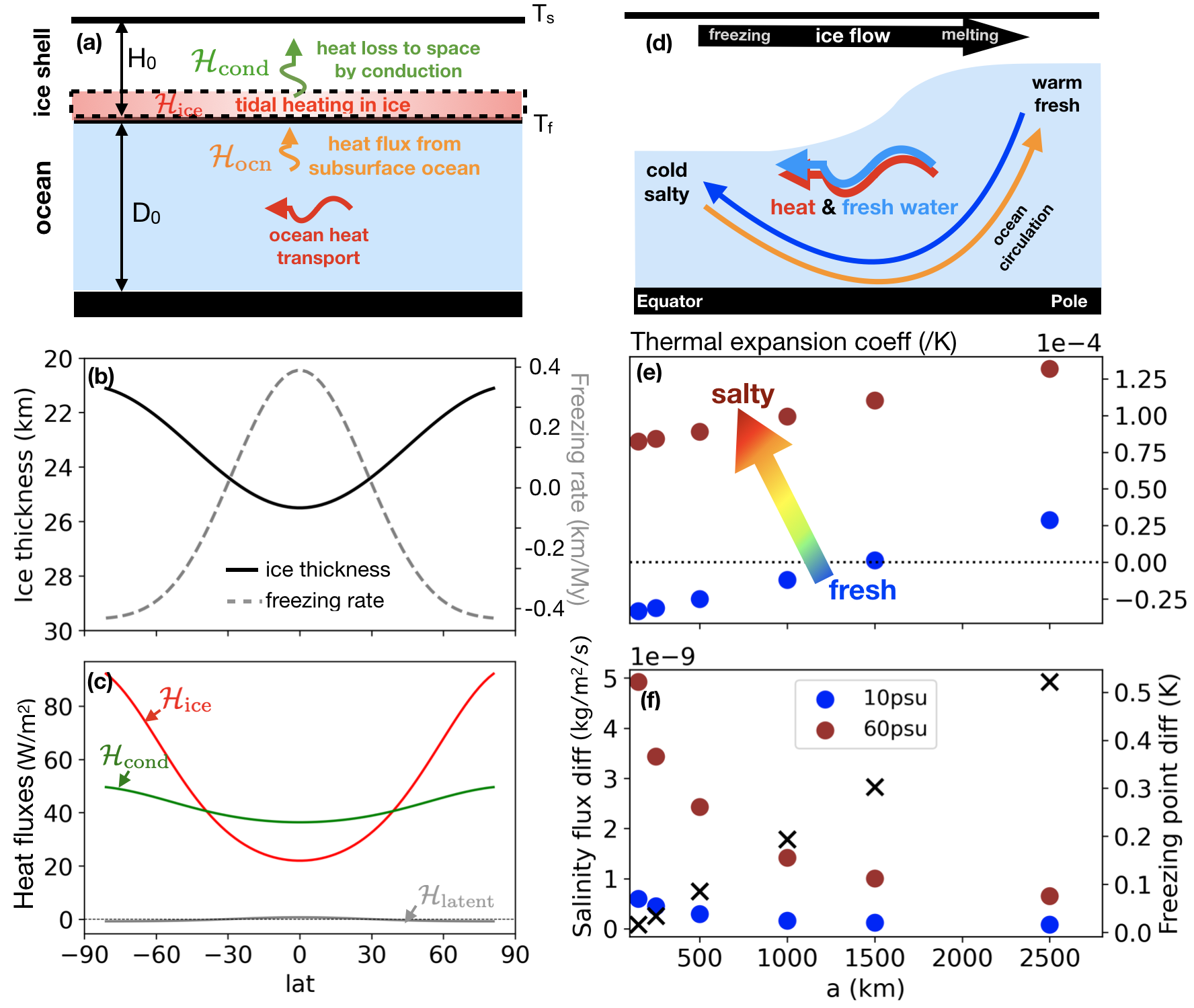}
    \caption{\small{Panel (a) sketches the primary sources of heat and heat fluxes, which include: heating due to tidal dissipation in the ice $\mathcal{H}_{\mathrm{ice}}$, the heat flux from the ocean to the ice $\mathcal{H}_{\mathrm{ocn}}$ and the conductive heat loss to space $\mathcal{H}_{\mathrm{cond}}$. Ocean heat transport is shown by the horizontal arrow. Panel (b) shows the default ice shell thickness profile considered here a black solid curve, which is thinner over the poles because ice dissipation amplifies going poleward \citep{Beuthe-2019:enceladuss}. The gray dashed curve shows the freezing (positive) and melting rate (negative) required to maintain a steady state based on an upside-down shallow ice flow model (see appendix for details). In this calculation, the default 2500~km radius is considered. Panel (c) shows the profiles of $\mathcal{H}_{\mathrm{ice}}$, $\mathcal{H}_{\mathrm{cond}}$ and $\mathcal{H}_{\mathrm{latent}}$ given the information in panel (b). Panel (d) sketches the key physical processes in an ocean covered by an ice shell with varying thickness (see main text for description). Panel (e) shows how thermal expansion coefficient under the ice shell varies with the satellite's size (gravity), at 10~psu (blue) and 60~psu (brown) ocean salinities. Panel (f) shows the salinity forcing (equatorial minus polar salinity flux, dots) and the temperature forcing (the freezing point difference under the equatorial and polar ice shell, crosses) as a function of the moon's radius.}}
    \label{fig:EOS-Hice-Heatflux}
  \end{figure*}

  The system considered by this study is sketched in Fig.~\ref{fig:EOS-Hice-Heatflux} -- a 56-km deep ocean covered by an ice shell that is about 20~km thick. A poleward-thinning ice shell is assumed, given the fact that the tidal dissipation in the ice shell amplifies over the poles \citep{Beuthe-2019:enceladuss}. 
  \begin{equation}
    \label{eq:ice-thickness}
    H(\phi)=H_0+\Delta H P_2(\sin\phi),
  \end{equation}
   where $H_0=20$~km is the mean ice thickness, $P_2$ is the 2nd order Legendre polynomials, and $\phi$ denotes latitude. Here, the ice thickness variation is assumed to follow the $P_2$ profile for simplification, and unless otherwise mentioned, the equator-to-pole thickness difference $\Delta H$ is set to $3$~km. This default ice thickness profile is shown by the solid curve in Fig.~\ref{fig:EOS-Hice-Heatflux}b. One would expect ice shell to be thinner over the poles as can be seen in Fig.~\ref{fig:EOS-Hice-Heatflux}b, because the tidal dissipation produced in the ice is stronger there \citep{Beuthe-2018:enceladuss, Beuthe-2019:enceladuss, Kang-Flierl-2020:spontaneous}. 
   In a situation where the ice shell is thinner over the equator ($\Delta H<0$), the results here can still apply after reversing the sign of the circulation and heat transport. Also, we expect the qualitative results obtained in this work to hold when the ice thickness variation follows a profile other than $P_2$, as long as the ice thickness is relatively simple (poleward-thinning or poleward-thickening) and is symmetric about the equator.

   With ice flowing from thick ice regions to thin ice regions \footnote{Ice convection is not considered here. The ice flow model used here assumes a purely conductive ice shell, and if the ice shell is instead convective, results may change.}, this equator-to-pole ice thickness gradient won't last, unless freezing/melting can constantly enhance the poleward-thinning ice topography. The ice freezing/melting, in turn, is governed by the ice shell heat budget. Driven by the hundreds of degree temperature difference between the water-ice interface and the ice surface, the ice constantly loses heat in form of heat conduction. On Enceladus and Europa, where the ice shell is likely around 20~km thick, the heat loss rate $\mathcal{H}_{\mathrm{cond}}$ is around 40~mW/m$^2$ on global average, and is faster over regions where ice is thin and over the poles where the ice surface temperature is low, as shown by the green curve in Fig.~\ref{fig:EOS-Hice-Heatflux}c. To balance the heat loss, the ice shell and the silicate core needs to produce heat, however, the core-shell heat partition is poorly understood \citep{Beuthe-2019:enceladuss, Choblet-Tobie-Sotin-et-al-2017:powering}. In this work, I will focus on the shell-heating scenario and discuss the potential impacts of bottom heating only toward the end. The ice heat production due to tidal flexing (denoted by $\mathcal{H}_{\mathrm{ice}}$) amplifies over the polar regions even if the ice is completely flat \citep{Beuthe-2018:enceladuss}, and this polar-amplifying pattern is further enhanced by the poleward thinning ice geometry through the ice rheology feedback, which concentrates heat production toward regions with thinner ice shell and hence weaker mechanical strength \citep{Beuthe-2019:enceladuss}. Shown by a red curve in Fig.~\ref{fig:EOS-Hice-Heatflux}c is $\mathcal{H}_{\mathrm{ice}}$ with the default ice thickness profile (Eq.~\ref{eq:ice-thickness}). Besides, the equatorial freezing and polar melting required to maintain the prescribed ice thickness gives rise to almost negligible latent heat release $\mathcal{H}_{\mathrm{latent}}$, shown by a gray curve in Fig.~\ref{fig:EOS-Hice-Heatflux}c. At the bottom of the ice, there may also be heat exchange with the ocean (denoted by $\mathcal{H}_{\mathrm{ocn}}$), manifested by ocean circulation and heat transport. In an equilibrium state, the ice shell heat budget needs to be in balance, which states as the vanishing of the sum of all the heating terms.

   From ocean's perspective, however, it is not the heat budget of the ice shell that matters, but the heat and salinity fluxes from the ice. These fluxes can be derived, as long as the ice geometry is given. In direct contact with ice, the ocean temperature at the water-ice interface will be relaxed toward the local freezing point, which is lower under a thick ice shell because of the high pressure (see Eq.~\ref{eq:freezing-point} in the appendix). Also, assuming ice shell is in mass equilibrium, equatorial freezing and polar melting is required to prevent the ice shell from being flattened by the pressure-driven ice flow (see the dashed gray curve in Fig.~\ref{fig:EOS-Hice-Heatflux}b). The freezing/melting will then induce salinity exchange with the subsurface ocean. Under these forcings, water over the poles become warmer and fresher than the water at low latitudes. The resultant density variations drive ocean circulation and eddies, transporting heat down-gradient from the poles to the equator, which in turns affect the heat budget of the ice shell, leading to changes in the ice geometry. Therefore, the critical task for predicting the ice shell geometry is to determine the dependency of the ocean heat transport (OHT) on the ice shell geometry via their heat and salinity fluxes under various satellite parameters (such as ocean salinity, gravity, size etc.).

   In section~\ref{sec:scaling-law-OHT} and section~\ref{sec:numerical-scaling-test}, I will derive scaling laws for the dependency of OHT on the ice shell geometry under various satellite parameters, test them against numerical simulations. Since we already know how the other heating terms (conductive heat loss, ice dissipation and latent heating) depends on $\Delta H$, this generic formula for OHT as a function of $\Delta H$ and the satellite parameters (such as ocean salinity, gravity, size etc.) allows one to solve for the equilibrium $\Delta H$ for a specific icy satellite under the condition that the ice shell heat terms should balance one-another. This is done in section~\ref{sec:equilibrium-dH}. It should be noted that, throughout this work, the heating produced in the silicate core is assumed to be zero and all heat is assumed to be generated in the ice shell. The impacts of core heating is discussed in the conclusion.

   More details about the ocean circulation model, ice flow model, and the tidal dissipation model can be found in the appendix.


\section{Why size matters?}
\label{sec:size-effect}   

  Using parameters relevant for Enceladus, \cite{Kang-Mittal-Bire-et-al-2021:how} show that the circulation that arises from the surface heat and freshwater forcing can go either direction depending on the ocean salinity: in the low-salinity limit, temperature-induced density variation dominates, and the warm polar water would sink as sketched by the blue arrow in Fig.~\ref{fig:EOS-Hice-Heatflux}d because fresh water contracts upon warming (anomalous expansion); whilst in the high-salinity limit, the anomalous expansion is suppressed, and both salinity- and temperature-induced density gradients contribute to downwelling at low-latitudes, as sketched by the orange arrow in Fig.~\ref{fig:EOS-Hice-Heatflux}d. 

  When considering icy satellites larger than Enceladus (most of the icy satellites of interest are), the following changes are expected:
  \begin{itemize}
\item The thermal expansion coefficient will become more positive, and eventually anomalous expansion will be completely suppressed even if the ocean is relatively fresh. Shown in Fig.~\ref{fig:EOS-Hice-Heatflux}e are the dependence of thermal expansion coefficient under the ice shell on the icy moon's radius, $a$, for two different salinities, 10~psu and 60~psu. Anomalous expansion doesn't occur in an ocean with 60~psu salinity regardless of the size of the satellite, yet it does occur in a 10-psu ocean, but only when $a< 1500$~km \footnote{20~km ice shell is assumed here.}, approximately the size of Europa. Therefore, with large enough planetary size, downwelling will only occur over the equator regardless of the ocean salinity.
\item The temperature difference under the ice shell between the equator and the pole $\Delta T$ will increase, as shown by the cross markers in Fig.~\ref{fig:EOS-Hice-Heatflux}f. The ocean temperature adjacent to the ice shell should be close to the local freezing point (denoted by $T_f$), and $T_f$ decreases with pressure linearly. 
  \begin{equation}
    \label{eq:deltaT}
     \Delta T=\Delta T_f=b_0\Delta P=b_0\rho_i g_0 \Delta H,
   \end{equation}
   where $\Delta H$ is the prescribe ice thickness difference between the equator and the poles, $g_0$ is the moon's surface gravity, $b_0=-7.61\times10^{-4}$~K/dbar and $\rho_i$ is the ice density. With $\Delta H$ fixed, $\Delta T \propto g \propto a$.
 \item  Salinity forcing will weaken. The blue and brown dots in Fig.~\ref{fig:EOS-Hice-Heatflux}f show the salinity flux (mean salinity $S_0$ times the freezing rate $q$) difference between the equator and the poles for an ocean with mean salinity of 10~psu and 60~psu, respectively. The freezing rate $q$ is set to balance the divergence of ice flow. As derived in the appendix~A4, ice flow behaves like diffusion, and the flow divergence/convergence is proportional to $\Delta P$ divided by the distance square $a^2$, so
   \begin{equation}
     \label{eq:qscaling}
     q \propto \Delta P/a^2\propto a^{-1}.
   \end{equation}
  \item The same density gradient will drive a stronger ocean circulation and heat transport as a result of the stronger gravity -- fixing the bulk density of the satellite, surface gravity $g_0\propto a$. The stronger heat transport will then flatten the ice shell more efficiently. 
\end{itemize}

Given the above reasoning, one can see that, on larger icy moons, the OHT is likely to be 1) more efficient in flattening the ice shell and 2) dominantly controlled by temperature variations. Assuming the ice thickness varies by 30\% meridionally and a melting-point ice viscosity of $10^{14}$~Pa$\cdot$s, a back-of-the-envelope calculation will show that the salinity-induced density anomaly is only comparable to the temperature-induced one if the moon's size is smaller or comparable to that of Enceladus. In \citet{Kang-Jansen-2022:in}, the authors have demonstrated these points in a zonally symmetric framework, ignoring eddy transports. Their results show that the meridional OHT should scale with the moon's radius $a$, the equator-to-pole ice thickness difference $\Delta H$ and the Coriolis coefficient $f$ as follows,
\begin{equation}
  \label{eq:scaling-kappalim-2D}
 \mathcal{F}_{\mathrm{ocn}}\propto a^3\Delta H^{3/2}f^{-1} 
\end{equation}
when the circulation depth is limited by vertical diffusion, or
\begin{equation}
  \label{eq:scaling-Dlim-2D}
 \mathcal{F}_{\mathrm{ocn}}\propto a^3\Delta H^{2}f^{-2}, 
\end{equation}
when the circulation reaches the seafloor.

\section{Scaling laws for ocean heat transport (OHT) considering eddies.}
\label{sec:scaling-law-OHT}
Given that salinity forcing tends to be dominated by temperature forcing unless the size of the icy moon is comparable or smaller than Enceladus (see the previous section), only the temperature-induced density anomalies is considered here, following \citet{Kang-Jansen-2022:in}. For smaller icy satellites, the salinity-driven circulation may add to or cancel out the temperature-driven circulation depending on the ocean salinity, and as a result, the OHT could be off by one order of magnitude \citep{Kang-Mittal-Bire-et-al-2021:how}.

The eddy heat transport $\mathcal{F}_e$ can be represented by an equivalent diffusive process,
\begin{equation}
  \label{eq:heat-transport-diffusivity}
  \mathcal{F}_e=C_p\rho_0\kappa_e\partial_yT \cdot (2\pi  ad)\sim 2\pi C_p\rho_0\kappa_e\Delta T d,
\end{equation}
where $\kappa_e$ is the equivalent diffusivity of baroclinic eddies and $d$ denotes the depth that the density anomaly penetrates downward from the surface. The corresponding residual circulation can be written as
\begin{equation}
  \label{eq:residual-psi-xi-eq-1}
  \Psi^\dagger\sim (2\pi a) \rho_0  \kappa_e\frac{T_y}{T_z}\sim 2\pi\rho_0 \kappa_e d (\Delta T/\Delta_v T),
\end{equation}
where $\Delta_vT$ is the vertical temperature contrast across the depth $d$.

According to \citet{Held-Larichev-1996:scaling}, $\kappa_e$ can be estimated by
\begin{equation}
  \label{eq:held-larichev}
  \kappa_e=kVL_\beta,
\end{equation}
where $k=0.25$ is a constant, $V$ is the rms eddy velocity and $L_\beta$ denotes wavelength of the energy-containing eddies, which follows the Rhines scale
\begin{equation}
  \label{eq:rhine-scale}
  L_\beta\sim\sqrt{V/\beta}.
\end{equation}
This energy-containing wavelength is typically larger than the deformation radius $L_d$, the scale at which baroclinic instability happens:
\begin{equation}
  \label{eq:deformation-radius}
  L_d\sim \frac{Nd}{f}\sim \frac{\sqrt{\alpha g\Delta_v Td}}{f},
\end{equation}
 $f$ denotes Coriolis coefficient, and $N=\sqrt{g\rho_z/\rho_0}$ denotes the Brunt-Vasala frequency. Here, the vertical density gradient $\rho_z$ is estimated by $\rho_0\alpha\Delta_v T$ divided by $d$.

According to the scaling proposed by \citet{Held-Larichev-1996:scaling}, the ratio $L_\beta/L_d$ and $V/U$ ($U$ denotes the zonal jet speed) is governed by the supercriticality $\xi$,
\begin{equation}
  \label{eq:xi}
  \frac{L_\beta}{L_d}\sim \frac{V}{U}\sim \xi\equiv\frac{\rho_yf/\beta}{\rho_zd}\sim \frac{\Delta T}{\Delta_v T},
\end{equation}
where $\beta\sim f/a$ is the meridional gradient of $f$. In the above equation, $d$ denotes the depth, to which circulation and dynamics can penetrate.

The thermal wind balance connects $U$ with $d$,
  \begin{equation}
  \label{eq:thermal-wind}
  U\sim \frac{\alpha g \Delta T d}{f a}.
\end{equation}
A second constraint can be provided by the balance of vertical heat transport. In an equilibrium state, the horizontally integrated vertical diffusion of heat $\rho C_p\kappa_v(\Delta_vT/d)(2\pi a^2)$ should be equal to the downward heat transport by residual circulation $C_p\Psi^\dagger \Delta T$ \citep{Jansen-Ferrari-2013:equilibration} and that yields
\begin{equation}
  \label{eq:vertical-heat-balance}
  \kappa_va^2=\kappa_ed^2 \xi^2.
\end{equation}

\paragraph{$\kappa_v$-limit. }
  I first consider the $\kappa_v$-limited situation, where, because of the low vertical diffusivity, the densest isentrope initiated from the equator bends over and reaches the poles without intersecting with the bottom. In this scenario, $d<D$, $\Delta_vT=\Delta T$, and $\xi\sim 1$. As a result, $V$ and $U$, $L_d$ and $L_\beta$ are interchangeable \footnote{It is easy to verify that the thermal wind balance is consistent with $V\sim U$, $L_d\sim L_\beta$ and $\xi=1$.}, so $\kappa_e$ can be written as
  \begin{equation}
  \label{eq:kappae-xi-eq-1}
  \kappa_e=kUL_d=\frac{k\beta}{f^3}(\alpha g \Delta T d)^{3/2}\sim\frac{k}{af^2}(\alpha g \Delta T d)^{3/2}.
\end{equation}

The depth of density variations $d$ is constrained by vertical diffusivity $\kappa_v$ through Eq.~\eqref{eq:vertical-heat-balance}, leading to
\begin{equation}
  \label{eq:depth-xi-eq-1}
  d
  \sim \left(\frac{\kappa_v a^3f^2}{k}\right)^{2/7}\cdot\left(\alpha g\Delta T\right)^{-3/7} \propto f^{4/7}\Delta H^{-3/7},
\end{equation}
which indicates that the depth of density variation is insensitive to the moon size, but increases with the rotation rate and decreases with ice thickness gradients.
To make sure that the above solution is consistent with the assumptions, one needs to check whether $d$ is indeed smaller than the ocean depth $D$. If this is not satisfied, the system should be in the $D$-limit instead. If $d$ indeed turns out smaller than $D$, then we can substitute Eq.~\eqref{eq:depth-xi-eq-1} and Eq.~\eqref{eq:kappae-xi-eq-1} into Eq.~\eqref{eq:heat-transport-diffusivity}, which gives
\begin{equation}
  \label{eq:heat-transport-xi-eq-1}
  \mathcal{F}_{\kappa_v}
  \sim 2\pi \rho C_p \Delta T \left(\frac{k}{af^2}\right)^{2/7}(\alpha g\Delta T)^{3/7}(\kappa_va^2)^{5/7}\propto a^3\Delta H^{10/7}f^{-4/7}.
\end{equation}
To obtain the proportionality, the facts that $\Delta T\propto \Delta P \propto g\Delta H\propto a \Delta H$, $g\propto a$ and $\beta\propto f/a$ are used. The dependence of $\mathcal{F}_{\kappa_v}$ on $a$, $\Delta H$ are very similar to that of the $\kappa_v$-limited overturning circulation in the zonally symmetric case given by \citet{Kang-Jansen-2022:in} (see Eq.~\ref{eq:scaling-kappalim-2D}).

Scaling laws can also be obtained for $L_d$ and $L_\beta$, which should characterize the size of eddies and the width of jets,
\begin{equation}
  L_{\beta,\kappa_v}=L_{d,\kappa_v}=(\alpha g\Delta T)^{2/7}(\kappa_v a^3/k)^{1/7}f^{-5/7}\propto a \Delta H^{2/7} f^{-5/7} \label{eq:deformation-rhines-xi-eq-1}
\end{equation}
and for the jet speed $U$
\begin{equation}
  U_{\kappa_v}=\beta L_{\beta,\kappa_v}^2\sim (\alpha g\Delta T)^{4/7}(\kappa_v /k)^{2/7}f^{-3/7}a^{-1/7}\propto a\Delta H^{4/7}f^{-3/7}. \label{eq:jet-xi-eq-1}
\end{equation}
In the above equations, subscript $\kappa_v$ indicates that this is the $\kappa_v$-limit scaling. Noticeably, the above scaling laws predict eddy size and jet width to grow linearly with moon's radius $a$, meaning the dominant wavenumber and the number of jets will be insensitive to the moon's radius, but the jets will be stronger ($U\propto a$). Besides, slower rotation and stronger ice thickness gradient will make the eddies and jet stronger and larger in size.

\paragraph{$D$-limit. }
For those scenarios, where isentropes intersect with the seafloor, $d=D$, $\Delta_vT<\Delta T$, and $\xi>1$ (supercritical). The equivalent eddy diffusivity should then be written as
\begin{equation}
  \label{eq:kappae-xi-gt-1}
  \kappa_e=kVL_\beta=kU\sqrt{U/\beta}\xi^{3/2}=\frac{k\beta}{f^3}(\alpha g \Delta T D)^{3/2}\xi^{3/2}.
\end{equation}
From the above equation and the vertical heat balance (Eq.~\ref{eq:vertical-heat-balance}), $\xi$ can be solved
\begin{equation}
  \label{eq:xi-gt-1}
  \xi=\left(\kappa_v a^2\frac{f^3}{k\beta}\right)^{2/7}(\alpha g \Delta T)^{-3/7}D^{-1}.
\end{equation}
Substituting the Eq.~\eqref{eq:kappae-xi-gt-1} and Eq.~\eqref{eq:xi-gt-1} into Eq.~\eqref{eq:heat-transport-diffusivity} yields the eddy heat transport
\begin{equation}
  \label{eq:heat-transport-xi-gt-1}
  \mathcal{F}_{D}\sim 2\pi \rho C_pD \Delta T \left(\frac{k\beta}{f^3}\right)^{4/7} (\alpha g\Delta T)^{6/7} (\kappa_va^2)^{3/7} \propto a^3\Delta H^{13/7}f^{-8/7}.
\end{equation}
The above scaling exhibits a stronger dependence on $\Delta H$ and $f$ than the $\xi=1$ scaling, and it becomes dependent on the ocean depth $D$ because the vertical diffusion now is strong enough to communicate the bottom and the ice shell. Again, this scaling is comparable to the 2D scaling obtained by \citet{Kang-Jansen-2022:in} (see Eq.~\ref{eq:scaling-Dlim-2D}).

By assumption, the penetration depth of the T/S anomalies is the entire ocean depth $D$, i.e.,
\begin{equation}
  \label{eq:depth-xi-gt-1}
  d=D.
\end{equation}
Substituting Eq.\ref{eq:depth-xi-gt-1} and Eq.\ref{eq:xi-gt-1} into the definitions of deformation radius and Rhines scale (Eq.~\ref{eq:rhine-scale}, Eq.~\eqref{eq:xi}, Eq.~\ref{eq:deformation-radius}), the following scalings are obtained:
\begin{eqnarray}
 L_{d,D}&=&(\alpha g\Delta T)^{5/7}(\kappa_va^2/(k\beta))^{-1/7}Df^{-10/7}\propto a^{9/7}\Delta H^{5/7}f^{-11/7}  \label{eq:deformation-xi-gt-1}\\
  L_{\beta,D}&=& (\alpha g\Delta T)^{2/7}(\kappa_va^2/(k\beta))^{1/7}f^{-4/7}\propto a\Delta H^{2/7} f^{-5/7}  \label{eq:rhine-xi-gt-1}\\
  U_{D} &=& (\alpha g\Delta T) D f^{-2}\beta\propto a\Delta H f^{-1}. \label{eq:U-xi-gt-1}          
\end{eqnarray}

Just as in the $\kappa_v$-limit scaling, the dominant eddy wavenumber and jet number should be insensitive to the radius of the moon since the power of the $a$ factors for $L_{d,D}$ and $L_{\beta,D}$ are both equal to one. Also, slower rotation and stronger ice thickness gradient will enhance the jet speed and increase the sizes of jets and eddies.

Combining the two scalings together, the final OHT should be equal to the lower value between the two scalings given by Eq.~\eqref{eq:heat-transport-xi-eq-1} and Eq.~\eqref{eq:heat-transport-xi-gt-1}, i.e.,
\begin{equation}
  \label{eq:heat-transport}
  \mathcal{F}_{\mathrm{ocn}}\sim \min\left\{\mathcal{F}_{\kappa_v},\mathcal{F}_{D}\right\},
\end{equation}
because both ocean depth and vertical diffusivity constrain how much heat can be transported by the ocean.

  \begin{figure*}[htbp!]
    \centering
    \includegraphics[width=\textwidth]{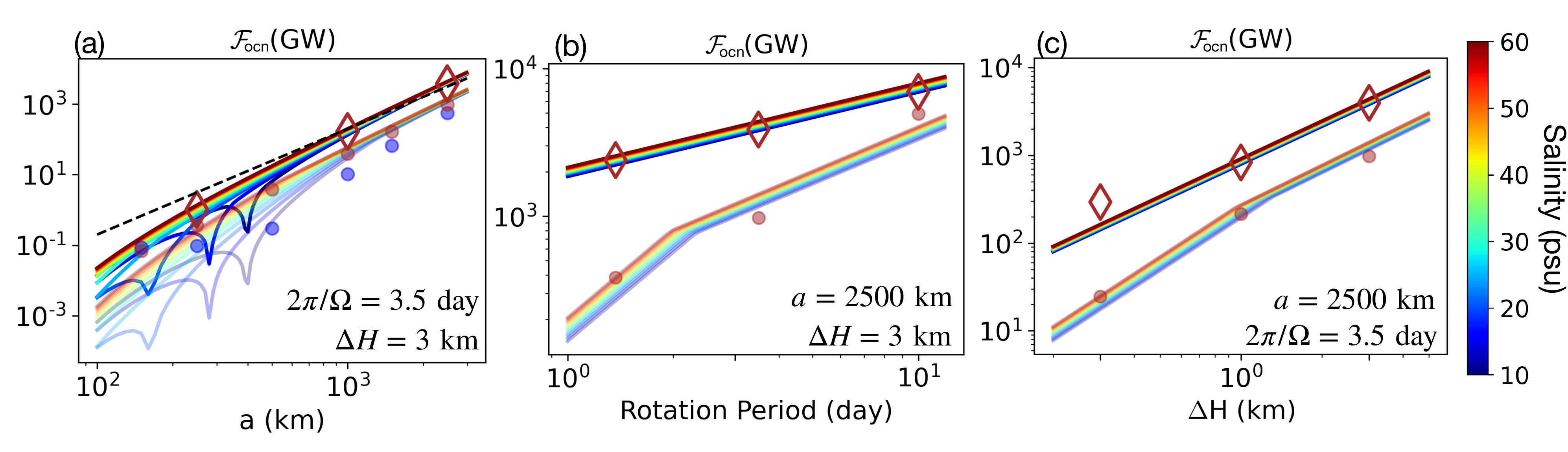}
    \caption{\small{Numerical verification of ocean heat transport (OHT) scalings. From panel (a) to (c) bottom shows the dependence on the moon's size $a$, the rotation rate reflected by the Coriolis coefficient $f$, and on the equator-to-pole ice thickness difference $\Delta H$. The lines in highly saturated colors present the 3D scaling given by Eq.~\eqref{eq:heat-transport-xi-eq-1} and Eq.~\eqref{eq:heat-transport-xi-gt-1}, and lines in lighter colors present the 2D scaling given by \citet{Kang-Jansen-2022:in}. Scattered on top are the diagnosed OHT from 3D numerical experiments (diamond markers) and 2D numerical experiments (dots). Different colors are used to differentiate different ocean salinities: from blueish color to reddish color, salinity increases. Default parameters used in the scaling and the numerical experiments can be found in Table~\ref{tab:parameters}. }}
    \label{fig:eddy-scaling}
  \end{figure*}

  \section{Examine the theory using numerical simulations.}
  \label{sec:numerical-scaling-test}
  Shown in Fig.~\ref{fig:eddy-scaling}(a-c) is the predicted $\mathcal{F}_{\mathrm{ocn}}$ as a function of the icy moon's size, rotation rate and equator-to-pole thickness difference in highly saturated colors. Blueish colors denote fresh ocean and reddish colors denote salty ocean. Default parameters can be found in Table~\ref{tab:parameters}. For comparison, the corresponding 2D scalings given by \citet{Kang-Jansen-2022:in} are shown in lighter colors. To verify these scaling laws, I ran 2D and 3D numerical simulations varying the three parameters ($a$, $f$ and $\Delta H$) and diagnose the OHT from the model output. The 2D and 3D results are marked on Fig.~\ref{fig:eddy-scaling} using dots and diamond markers, respectively. They match the prediction (Eq.~\ref{eq:heat-transport}) up to a factor of 2. The matching seems particularly good in panel (a), because the range of $\mathcal{F}_{\mathrm{ocn}}$ there is much wider so that the offset appears smaller in comparison.

    \begin{figure*}[htbp!]
    \centering
    \includegraphics[width=\textwidth]{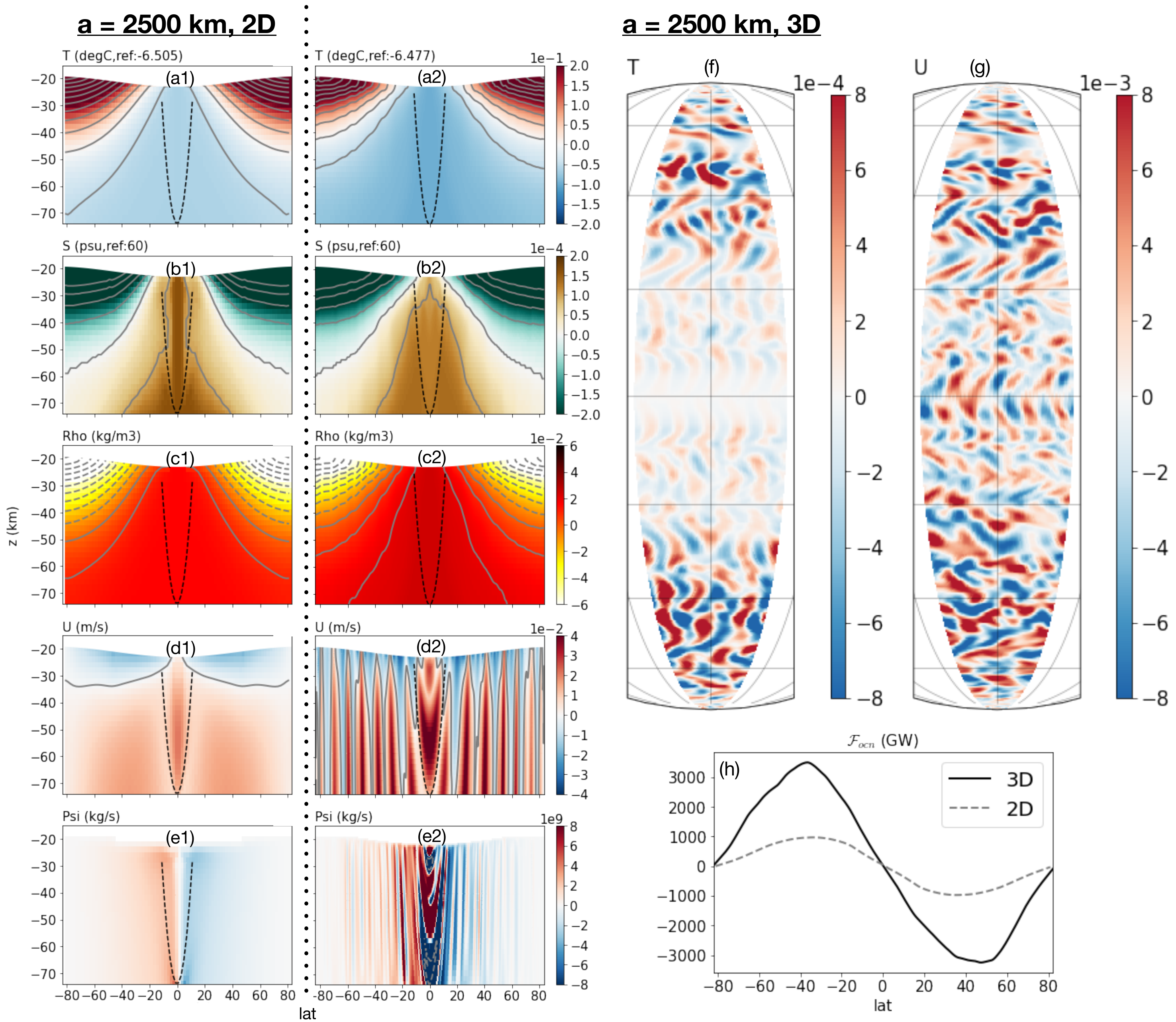}
    \caption{\small{Ocean circulation and thermodynamic state under 2D and 3D configurations. The left column (panels a1-e1) shows temperature, salinity, density, zonal flow and meridional streamfunction from a zonally symmetric 2D simulation. The second column (panels a2-e2) shows the same thing for the 3D simulation. Panel (f) and panel (g) present the temperature and zonal flow anomalies from the zonal mean in a plane view. Panel (h) presents the vertically and zonally integrated meridional ocean heat transport diagnosed from the 2D (dashed) and 3D model (solid). In this defualt setup: moon's radius $a=2500$~km, ocean salinity $S=60$~psu, equator-to-pole ice thickness difference $\Delta H=3$~km, and Europa's rotation period (3.5~day). }}
    \label{fig:3d-dynamics}
  \end{figure*}

  The OHT is governed by the ocean's dynamic and thermodynamic states. Presented in Fig.\ref{fig:3d-dynamics} are the model solutions obtained under 2D and 3D configurations. In these simulations, moon's radius $a=2500$~km, ocean salinity $S=60$~psu, equator-to-pole ice thickness difference $\Delta H=3$~km, and Europa's rotation period (3.5~day) is used. Due to the pressure gradient induced by the poleward thinning ice geometry, the freezing point under the ice is higher over the poles compared to the equator, driving the poleward warming pattern seen in Fig.\ref{fig:3d-dynamics}-a. Meanwhile, in order to sustain the ice geometry against the flattening due to ice flow (Eq.\ref{eq:ice-flow}), equatorial ice needs to freeze and polar ice shell needs to melt. This in turn drives a meridional salinity gradient seen in Fig.\ref{fig:3d-dynamics}-b. High ocean salinity and high pressure suppress the anomalous expansion behavior (contract upon warming) near the freezing point, which typically happens on small icy moon with fresh ocean. As a result, temperature and salinity anomalies both contribute to the high density in low latitudes, driving sinking motions there (see Fig.\ref{fig:3d-dynamics}-e). The circulation is forced to be mostly aligned with the direction of rotation in the interior. This is because any motions moving closer or away from the rotating axis will lead to eastward/westward acceleration by virtue of angular momentum conservation, and the resultant zonal flows will be too strong to be in thermal-wind balance with the weak density variation. Flows across the direction of the rotation axis concentrate near the two rough boundaries at the top and bottom, because only in appearance friction, radial flows can be sustained without inducing imbalance in the zonal momentum budget. In an equilibrium state, the upper part of the ocean flows westward and the lower part of the ocean flows eastward (Fig.\ref{fig:3d-dynamics}-d), in thermal wind balance with the density distribution (Fig.\ref{fig:3d-dynamics}-c). 
  
  The major differences between 3D and 2D configurations lie in the zonal flow field (Fig.\ref{fig:3d-dynamics}-d): there are jets formed in 3D due to the Renold stress associated with the baroclinic eddies, whose structure is presented in Fig.\ref{fig:3d-dynamics}-f,g. Due to the strong rotation effect, the eddy motions tend to be aligned with the direction of rotation axis, forming convective ``rolls'' along the equatorial plane and wave-like structures in higher latitudes, consistent with recent icy moon studies \cite{Ashkenazy-Tziperman-2021:dynamic, Soderlund-Schmidt-Wicht-et-al-2014:ocean, Bire-Kang-Ramadhan-et-al-2022:exploring, Kang-Bire-Campin-et-al-2020:differing}. These eddies facilitate stronger equatorward heat transport than the overturning circulation in 2D configuration, as shown by Fig.\ref{fig:3d-dynamics}h. It should be noted that some previous works have jets and Taylor columns even under 2D configuration \citep{Ashkenazy-Tziperman-2021:dynamic}. This difference arises from the different model configuration: in their work, the ocean is heated from below and no-slip boundary condition is applied to the top and the bottom; whereas in this work, the bottom heating is assumed to equal to zero and the boundary drag is parameterized by a linear drag toward zero.
  
  At different the moon's sizes, rotation rates and the ice thickness variations, the dynamics remain qualitatively the same, as shown by Fig.\ref{fig:3d-dynamics-a250}-Fig.\ref{fig:3d-dynamics-H03} in the appendix. On a smaller icy moon, the meridional temperature gradient weakens due to its weaker gravity (compare Fig.\ref{fig:3d-dynamics-a250}a and Fig.\ref{fig:3d-dynamics-a1000}a with Fig.\ref{fig:3d-dynamics}a). That in turn weakens the circulation (see panel e1) and eddies (see panel f,g), leading to a much weaker heat transport (see panel h) as suggested by Eq.\ref{eq:heat-transport-xi-eq-1} and Eq.\ref{eq:heat-transport-xi-gt-1}. According to Eq.~\eqref{eq:deformation-rhines-xi-eq-1} and Eq.~\eqref{eq:rhine-xi-gt-1}, the number of jets remain more of less unchanged -- this can be seen from the panel-d2 of Fig.\ref{fig:3d-dynamics-a250}, Fig.\ref{fig:3d-dynamics-a1000} and Fig.\ref{fig:3d-dynamics}.
  
  Fast rotation suppresses the circulation, eddies and thereby OHT, as suggested by both 2D scalings  (Eq.\ref{eq:scaling-kappalim-2D} and Eq.\ref{eq:scaling-Dlim-2D}) and 3D scalings (Eq.\ref{eq:heat-transport-xi-eq-1} and Eq.\ref{eq:heat-transport-xi-gt-1}). This trend is reflected by sensitivity tests shown in Fig.\ref{fig:3d-dynamics-O1.37} and Fig.\ref{fig:3d-dynamics-O10}, where a shorter rotation period of 1.37~days (Enceladus' rotation period) and a longer rotation period of 10~days are employed, respectively. Also, when rotation rate varies, the width of jets and eddies should change accordingly. Strong rotation leads to smaller deformation radius $L_d=(NH/f)$ (governing the eddy size, Eq.\ref{eq:deformation-radius}) and smaller Rhines scale $L_\beta=\sqrt{U/\beta}$ (governing the jet width, Eq.\ref{eq:rhine-scale}), consistent with panel~(d,f,g) of Fig.\ref{fig:3d-dynamics-O1.37}, Fig.\ref{fig:3d-dynamics} and Fig.\ref{fig:3d-dynamics-O10}.
  The dependence of OHT on the ice thickness variation $\Delta H$ is quite intuitive. When the ice is flatter, the temperature/salinity variations, eddy amplitude, eddy size and heat transport all decrease, as shown by Fig.\ref{fig:3d-dynamics-H1} and Fig.\ref{fig:3d-dynamics-H03}.

  As a further verification of the theory, the jet speed is diagnosed from the simulations by subtracting the averaged $U$ from the maximum zonal mean zonal flow, and is compared against the predictions given by Eq.~\eqref{eq:jet-xi-eq-1} and Eq.~\eqref{eq:U-xi-gt-1}. The averaged spacing between jets is identified by 'numpy.find\_peaks', and is compared against Eq.~\eqref{eq:deformation-rhines-xi-eq-1} and Eq.~\eqref{eq:rhine-xi-gt-1}. Finally, the prominent eddy size is computed by averaging different wavelength by its corresponding power spectrum of $U$ field between 40N/S and 60N/S, and the results are compared against Eq.~\eqref{eq:deformation-rhines-xi-eq-1} and Eq.~\eqref{eq:rhine-xi-gt-1}, because the eddies' energy-containing scale follows the Rhines scale \citep{Held-Larichev-1996:scaling}. All these diagnostics are carried out in high-latitudes inside the tangent cylinder -- a cylinder whose sides are parallel to the moon's axis of rotation and are tangential (hence the name) to the ocean's floor at the equator, because the dynamics outside the tangent cylinder are very different \citep{Kang-Bire-Campin-et-al-2020:differing, Bire-Kang-Ramadhan-et-al-2022:exploring}. As shown in Fig.\ref{fig:eddy-scaling-2}, the theory also catch the eddy and jet characteristics reasonably well, indicating that the agreement between the predicted and simulated $\mathcal{F}_{\mathrm{ocn}}$ may not be a coincidence.
  
  \section{The equilibrium equator-to-pole ice thickness gradient.}
  \label{sec:equilibrium-dH}

  \subsection{Scaling Theory}
  \label{sec:dH-scaling}
  The dependence of $\mathcal{F}_{\mathrm{ocn}}$ on orbital parameters and $\Delta H$ can be used to predict the equilibrium ice thickness variation using the fact that the ice shell heat budget should be closed. This analysis has been done by \citet{Kang-Jansen-2022:in} but using the 2D scaling for $\mathcal{F}_{\mathrm{ocn}}$. Here, I repeat the process for 3D scalings.

  First, I need to convert $\mathcal{F}_{\mathrm{ocn}}$ to heat flux anomaly received by the ice shell. Assuming that the heat transported from the polar regions to the equatorial regions by the ocean is evenly distributed over a half hemisphere with a surface area of $\pi a^2$, the heat flux per area received by the equatorial ice shell and the heat flux leaving the polar ice shell equal to
  \begin{widetext}
\begin{equation}
  \label{eq:Hocn}
  \mathcal{H}_{\mathrm{ocn}}=\left.\mathcal{F}_{\mathrm{ocn}}\right/(\pi a^2)\sim
  \begin{dcases}
    2 \rho C_p \Delta T \left(\frac{k}{a^3f^2}\right)^{2/7}(\alpha g\Delta T)^{3/7}\kappa_v^{5/7}\propto a\Delta H^{10/7}f^{-4/7} & \text{for }\kappa_v\text{-limit}\\
    2 \rho C_pD \Delta T \left(\frac{k\beta}{a^2f^3}\right)^{4/7} (\alpha g\Delta T)^{6/7} \kappa_v^{3/7} \propto a\Delta H^{13/7}f^{-8/7}& \text{for D-limit}
  \end{dcases}
\end{equation}
\end{widetext}

\begin{figure*}
    \centering \includegraphics[width=0.6\textwidth]{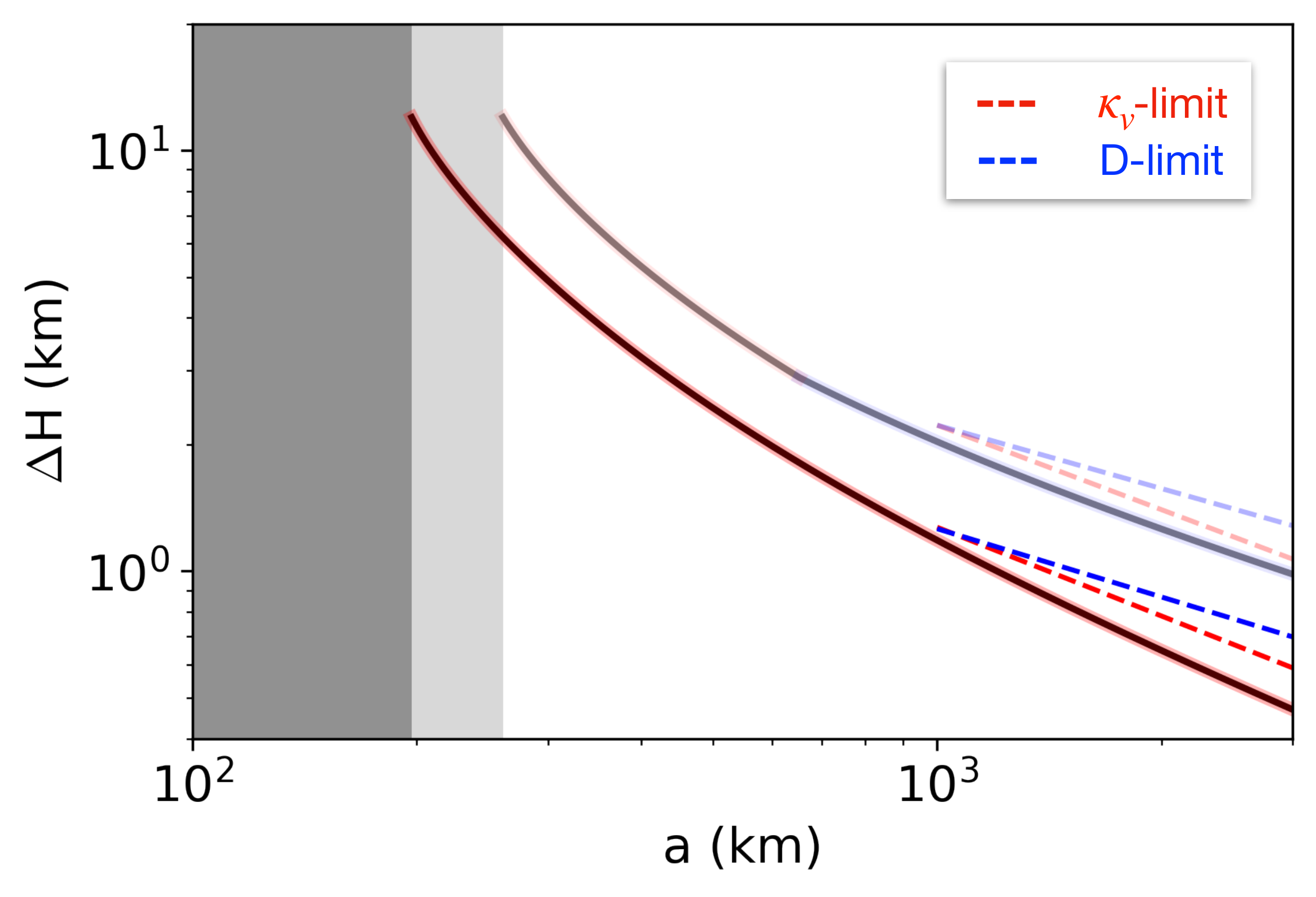}
    \caption{\small{Predicted equator-to-pole ice thickness difference in equilibrium solved from Eqs.~(\ref{eq:heat-balance}) and (\ref{eq:Hocn}). Red shading marks the $\kappa_v$-limit regime and blue shading marks the $D$-limit regime. Highly saturated colors represent 3D scalings and lighter colors represent 2D scalings obtained in \citet{Kang-Jansen-2022:in}. When $\Delta H>H_0/2$ ($H_0$ is the mean ice thickness), the polar ice shell thickness approaches zero, and the small $\Delta H$ assumption required for Eq.~\eqref{eq:heat-balance_small_dH} no longer holds. Those scenarios are considered to be run-away poleward-thinning and mask those with gray shadings.}}
    \label{fig:predicted-dH}
  \end{figure*}

In an equilibrium state, the ice shell needs to be in a heat balance everywhere, which means
\begin{equation}
  \label{eq:heat-budget}
  \mathcal{H}_{\mathrm{ice}}+\mathcal{H}_{\mathrm{latent}}+\mathcal{H}_{\mathrm{ocn}}=\mathcal{H}_{\mathrm{cond}}.
\end{equation}
$\mathcal{H}_{\mathrm{ice}}$ denotes the ice dissipation, $\mathcal{H}_{\mathrm{latent}}$ denotes the latent heat release and $\mathcal{H}_{\mathrm{cond}}$ denotes the conductive heat loss.
The latent heat release $\mathcal{H}_{\mathrm{latent}}=\rho_iL_fq$ tends to be small compared to the other terms, as can be seen from Fig.~\ref{fig:EOS-Hice-Heatflux}c, except when the moon size is very small and the ice thickness variation is very large. For simplicity, I drop $\mathcal{H}_{\mathrm{latent}}$. The remaining terms can all be written as a function of the ice topography $H$. $\mathcal{H}_{\mathrm{ice}}$ amplifies over regions with thinner ice due to the rheology feedback \citep{Beuthe-2019:enceladuss, Kang-Flierl-2020:spontaneous} following $\mathcal{H}_{\mathrm{ice}}(\phi)=\mathcal{H}_{\mathrm{ice0}}(\phi)\times(H_0/H(\phi))^2$, where $H_0$ is the mean ice thickness, $\mathcal{H}_{\mathrm{ice0}}(\phi)$ is the ice dissipation rate in a flat ice shell as a function of latitude $\phi$. The conductive heat loss $\mathcal{H}_{\mathrm{cond}}$ is inversely proportional to the local ice thickness $\mathcal{H}_{\mathrm{cond}}(\phi)=\mathcal{H}_{\mathrm{cond0}}\times(H_0/H(\phi))$. Since, the ice dissipation over the polar regions is roughly twice as strong as that over the equator in absence of ice topography \citep{Beuthe-2018:enceladuss}, I choose $\mathcal{H}_{\mathrm{ice0}}(\mathrm{pole})=1.25\overline{\mathcal{H}_{\mathrm{ice0}}}$ for the polar box and $\mathcal{H}_{\mathrm{ice0}}(\mathrm{eq})=0.75\overline{\mathcal{H}_{\mathrm{ice0}}}$ for the equatorial box, where $\overline{(\cdot)}$ denotes global mean. To guarantee global heat budget balance, $\overline{\mathcal{H}_{\mathrm{ice0}}}$ should be equal to $\mathcal{H}_{\mathrm{cond0}}\overset{\Delta}{=} \mathcal{H}$. Now, consider the equator-to-pole difference of the heat budget terms for an ice shell that has a mean thickness of $H_0$ and an equator-to-pole thickness difference of $\Delta H$ (equatorial ice shell is thicker). Keeping the first order terms, the heat budget simplifies to
\begin{widetext}
\begin{equation}
  \label{eq:heat-balance}
  2\mathcal{H}_{\mathrm{ocn}}=\Delta \mathcal{H}_{\mathrm{ice}}-\Delta \mathcal{H}_{\mathrm{cond}}\approx 1.25\mathcal{H}\left(1-\frac{\Delta H}{2H_0}\right)^{-2}-0.75\mathcal{H}\left(1+\frac{\Delta H}{2H_0}\right)^{-2}-\left[\mathcal{H}\left(1-\frac{\Delta H}{2H_0}\right)^{-1}-\mathcal{H}\left(1+\frac{\Delta H}{2H_0}\right)^{-1}\right].
\end{equation}
\end{widetext}
From Eqs.~(\ref{eq:heat-balance}) and (\ref{eq:Hocn}), $\Delta H$ can be solved numerically. The results are presented in Fig.~\ref{fig:predicted-dH}. The entire solution falls into the $\kappa_v$-limit regime, as denoted by the red shading (blue shading marks the $D$-limit regime).

If the ice thickness variation is small ($\Delta H/H_0\ll 1$), further simplification can be made to Eq.~\eqref{eq:heat-balance}. Taylor expanding Eq.~\eqref{eq:heat-balance} around small $\Delta H/H_0$ and dropping high order terms gives
\begin{equation}
  \label{eq:heat-balance_small_dH}
  2\mathcal{H}_{\mathrm{ocn}} \approx \frac{1}{2}\mathcal{H} 
\end{equation}
From Eqs. (\ref{eq:heat-balance_small_dH}) and (\ref{eq:Hocn}), $\Delta H$ can be solved analytically:
\begin{equation}
   \label{eq:limit-small-dH}
   \Delta H\approx
   \begin{dcases}
    \chi_\kappa \kappa_v^{-1/2}\alpha^{-3/10}f^{2/5}a^{-7/10},& \text{for }\kappa_v\text{-limit}\\
    \chi_D \kappa_v^{-3/13}\alpha^{-6/13}f^{8/13}a^{-7/13},& \text{for D-limit}\\
\end{dcases}
\end{equation}

where $\chi_\kappa=\left(\frac{\mathcal{H}}{8\rho C_p}\right)^{7/10}\left(4\pi G \rho_b/3\right)^{-13/10}k^{-1/5}(b\rho_i)^{-1}$ and $\chi_D=\left(\frac{\mathcal{H}}{8\rho C_pD}\right)^{7/13}\left(4\pi G \rho_b/3\right)^{-19/13}k^{-4/13}(b\rho_i)^{-1}$ are constants. To obtain the above solution, I have substituted $g$ with $4\pi G\rho_{\mathrm{bulk}}a/3$, where $\rho_{\mathrm{bulk}}=2500$~kg/m$^3$ is the bulk density.

From Eq.~(\ref{eq:limit-small-dH}), it can be seen that, when $\Delta H\ll H_0$, the equilibrium ice shell thickness variation $\Delta H$ should decrease with the moon's size and increases with the rotation frequency, following $\Delta H\propto f^{2/5}a^{-7/10}$ or $f^{8/13}a^{-7/13}$ depending on the dynamic regime of the ocean. The results here again qualitatively agree with the 2D scaling results, which follow $\Delta H\propto (f/a)^{2/3}$ or $f/a^{1/2}$ for $\kappa_v$-limit and $D$-limit, respectively, as shown by \citet{Kang-Jansen-2022:in}, except that the sensitivity to rotation rate is slightly lower here.

The asymptotic scalings (shown by blue and red dashed lines in Fig.~\ref{fig:predicted-dH}) provide a useful approximation to the full solution of Eq.~\eqref{eq:heat-balance} for relatively small to moderate $\Delta H$. For larger $\Delta H$, the sensitivity of $\Delta H$ on $a$ increases, and eventually, a run-away poleward-thinning happens when $a\approx 200$~km (masked by dark gray shading), due to the strengthening of the ice-rheology feedback.
Besides the dependence on $a$ and $f$, it can be seen from Eq.~(\ref{eq:limit-small-dH}) that a higher ocean salinity (leading to larger $\alpha$ and stronger salinity-driven circulation), a stronger turbulent diffusivity, and (provided sufficient turbulent mixing) a deeper ocean can also reduce $\Delta H$.

Compared to Enceladus, Europa is $6$ times larger in size, is rotating $3$ times slower, and its ocean is likely saltier \citep{Zolotov-2007:oceanic, zolotov2014can, Glein-Postberg-Vance-2018:geochemistry, Kang-Mittal-Bire-et-al-2021:how, Hand-Chyba-2007:empirical} and deeper \citep{Hand-Chyba-2007:empirical} -- all these differences suggest that Europa's ice shell may be much flatter than Enceladus'. Assuming $|\alpha|\sim 10^{-5}$-$10^{-4}$/K (Fig.~\ref{fig:EOS-Hice-Heatflux}e), $\kappa_v=10^{-3}$~m$^2$/s, $\gamma=10^{-4}$~m/s, Eq.\ref{eq:heat-balance} yield $\Delta H=8$~km for Enceladus \footnote{In reality, $\Delta H$ should be smaller if the following factors are account for: 1) the salinity-driven circulation, which could be crucial for small icy satellites and 2) $\mathcal{H}_{\mathrm{latent}}$ in the heat budget (Eq.\ref{eq:heat-budget}). The sensitivity of $\Delta H$ to orbital and ocean parameters dramatically increase when $\Delta H$ becomes comparable to the mean ice thickness $H_0$.}, in line with the strong ice topography in observations \citep{Iess-Stevenson-Parisi-et-al-2014:gravity, Beuthe-Rivoldini-Trinh-2016:enceladuss, Tajeddine-Soderlund-Thomas-et-al-2017:true, Cadek-Soucek-Behounkova-et-al-2019:long, Hemingway-Mittal-2019:enceladuss}. In contrast, using Europa parameters, $\Delta H$ is estimated to be only 0.7~km! This roughly matches the constraint based on limb profile measurements \citep{Nimmo-Thomas-Pappalardo-et-al-2007:global}. Finally, it should be noted that $\Delta H$ is sensitive to the vertical diffusivity/viscosity and is only valid under our assumptions (shell-heating, temperature-dominant density variation, Maxwell ice rheology etc.). Better understanding of the dissipative processes in the ocean driven by tides and libration motions \citep{Rekier-Trinh-Triana-et-al-2019:internal} is required to better constrain the ice shell geometry.

\subsection{Numerical Results for Enceladus and Europa}
 \label{sec:evolution}
 
 To demonstrate the potential impacts of the size of the icy satellite on its equilibrium ice shell geometry, I integrate an ice evolution model forward using Enceladus' and Europa's parameters, respectively. The model is modified based on \citet{Kang-Flierl-2020:spontaneous}. It calculates the melting induced by the tidal heating $\mathcal{H}_{\mathrm{ice}}$ (given by Eq.15 in the appendix), the down-gradient ice flow $\mathcal{Q}$ (given by Eq.13 in the appendix), the heat loss to space by conduction $\mathcal{H}_{\mathrm{cond}}$ (given by Eq.4 in the appendix), and heat transmitted upward by the ocean $\mathcal{H}_{\mathrm{ocn}}$, and evolves the ice thickness $H$ over time. The total thickness tendency can be symbolically expressed as 
\begin{equation}
  \label{eq:ice-evolution}
  \frac{dH}{dt}=\frac{\mathcal{H}_{\mathrm{cond}}(H)-\mathcal{H}_{\mathrm{ice}}(H)-\mathcal{H}_{\mathrm{ocn}}(H)}{L_f\rho_i } + \frac{1}{a\sin\phi}\partial_\phi \left(\sin\phi \mathcal{Q}(H)\right),
\end{equation}
where $L_f$ and $\rho_i$ are the latent heat of freezing and density for ice, $a$ is the moon's radius and $\phi$ denotes latitude. When the ice shell is thinner than 3~km, I assume that the ice shell will crack open under the tidal stress, and the resultant geysers will carry away large amounts of heat, preventing further melting. In the evolution model, I overwrite the thickness tendency with zero when $H<3$~km, to implicitly represent this additional heat loss.

The ocean-ice heat exchange $\mathcal{H}_{\mathrm{ocn}}$ is a new component that doesn't exist in \citet{Kang-Flierl-2020:spontaneous}. Inspired by the conceptual model, $\mathcal{H}_{\mathrm{ocn}}$ is parameterized as
\begin{equation}
  \label{eq:H-ocn}
  \mathcal{H}_{\mathrm{ocn}}=\mathrm{MIN}\left\{2 \rho C_p \Delta T \left(\frac{k}{a^3f^2}\right)^{2/7}(\alpha g\Delta T)^{3/7}\kappa_v^{5/7},\ 2 \rho C_pD \Delta T \left(\frac{k\beta}{a^2f^3}\right)^{4/7} (\alpha g\Delta T)^{6/7} \kappa_v^{3/7}\right\}
\end{equation}
where $H'$ is the deviation from the prescribed global mean ice thickness $H_0=20$~km and $\mathrm{MIN}\{\}$ selects whichever parameterization yields a smaller global standard deviation. A factor of 2 is multiplied to H' because the analysis before all use equator to pole thickness difference, which is twice as large as the equator/pole's deviation from the mean. Also, salinity-driven heat transport is assumed to be roughly comparable with the temperature-driven one, and that accounts for the other factor of 2 multiplied to the above formula.

\begin{figure*}
    \centering \includegraphics[width=\textwidth]{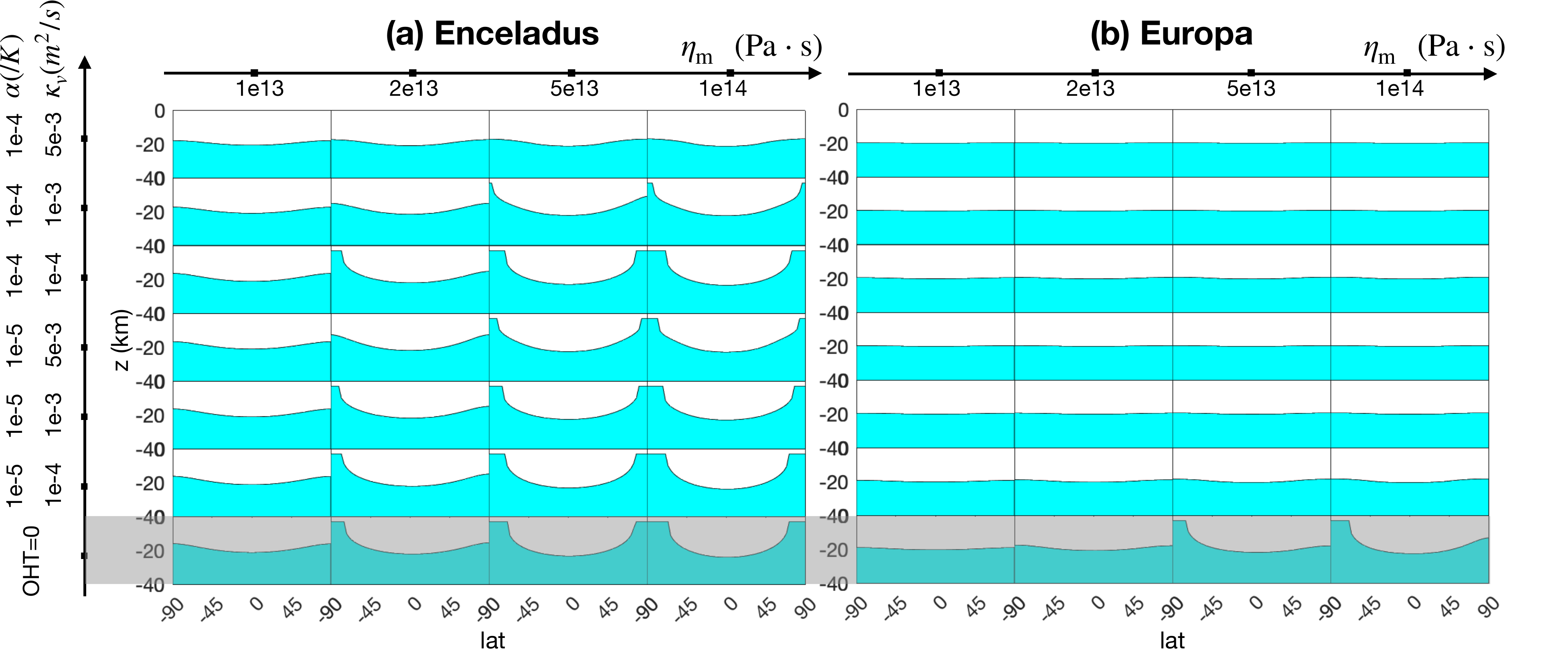}
    \caption{\small{Equilibrium ice shell geometries on Enceladus (panel a) and Europa (panel b) predicted by an ice evolution model (Eq.~\ref{eq:ice-evolution}) with parameterized ocean heat transport (Eq.~\ref{eq:H-ocn}). Blue color masks the ocean and white color masks the ice. 24 scenarios are considered for each moon, to account for the uncertainties associated with the ice shell rheology and the efficiency of ocean heat transport. The values for the three key parameters, $\eta_m$, $|\alpha|$ and $\kappa_v$, are shown on the left and upper sides. The top two rows have ocean heat transport amplified by another factor of 10 to represent the potential effect of eddies and other unforeseeable factors. The bottom row assumes zero ocean heat transport.}}
    \label{fig:ice-evolution}
  \end{figure*}

To account for the uncertainties associated with the ice shell rheology and the efficiency of ocean heat transport, a range of ice viscosities $\eta_m$ \footnote{Notice that a smaller $\eta_m$ means more freezing/melting is needed to counterbalance the ice flow, and thereby more latent heating and stronger salinity-driven circulation. I ignored these two effects in section~\ref{sec:dH-scaling}, which is why they do not depend on $\eta_m$ and can be thought of as representing the limit of large ice viscosity.}, $|\alpha|$ and $\kappa_v$ are considered for Enceladus and Europa. The equilibrium ice shell geometries are shown in Fig.~\ref{fig:ice-evolution}. For both moons, the equilibrium geometry also varies with the ice and ocean properties. The equilibrium ice shell tends to be flatter with smaller $\eta_m$ and higher $|\alpha|$ and $\kappa_v$.

The bottom row assumes no ocean heat transport, and significant ice thickness variations develop on both Enceladus and Europa. When the ice viscosity is not too low ($\eta_m>10^{-13}$~Pa$\cdot$s), ice is almost completely melt through over one or both poles due to the ice-rheology feedback. However, with ocean heat transport, the equilibrium ice geometry is largely flattened especially for Europa given its large size and slower rotation rate. The smoothing effect of ocean is particularly clearly shown in the right row, where ice flow is ineffective. If Europa's ocean is saltier than 50~psu as suggested by the strong magnetic induction signal \citep{Hand-Chyba-2007:empirical}, $\alpha$ should be closer to the upper bound, leading us to the conjecture that ice thickness variation on Europa is likely less than 1~km. When Enceladus' parameters are used instead, the ocean heat transport's impact on ice shell geometry is more limited. The equilibrium ice shell geometry obtained under the influence of OHT is similar to those without, unless the vertical diffusivity is very high. This sensitivity highlights the importance to understand the dissipative processes in the ocean driven by tides and libration motions \citep{Rekier-Trinh-Triana-et-al-2019:internal}. It should also be noted that the OHT scaling obtained in this work may not apply to scenarios with very strong ice thickness variation due to the changes of geometry and the nonlinear behavior of eddies at high amplitude. Among the 28 scenarios considered for Enceladus, five develop the hemispheric asymmetry seen in observation \citep{Iess-Stevenson-Parisi-et-al-2014:gravity, Hemingway-Mittal-2019:enceladuss}, suggesting that the symmetry-breaking mechanism proposed by \citet{Kang-Flierl-2020:spontaneous} could work in presence of ocean heat redistribution.

\section{Concluding remarks.}

The two scientific questions addressed here are 1) how the efficiency of the ocean heat transport (OHT) forced by the ice thickness variations varies with the icy moon's orbital parameters and 2) how the OHT in turn affects the equilibrium ice geometry. To do so, I derive scaling laws for the OHT on icy moons, inspired by previous theoretical work on the baroclinic eddies in the context of earth ocean or atmosphere \citep{Held-Larichev-1996:scaling, Karsten-Jones-Marshall-2002:role, Jansen-Ferrari-2013:equilibration}. These scaling laws are then verified by 3D general circulation simulations run for various planetary radii, rotation rates and associated ice thickness variations. It is found that heat convergence toward the thick-ice regions is more efficient on icy satellites with greater sizes and slower rotation rates. Therefore, those icy moons' ice shells are expected to be flatter.

Enceladus and Europa are two icy satellites in the solar system that are known to contain a global subsurface ocean \citep{Postberg-Kempf-Schmidt-et-al-2009:sodium, Thomas-Tajeddine-Tiscareno-et-al-2016:enceladus, Carr-Belton-Chapman-et-al-1998:evidence, Kivelson-Khurana-Russell-et-al-2000:galileo, Hand-Chyba-2007:empirical}. Despite their similar global-mean ice thickness and per-area heat production rate, Europa's ice shell is likely to undergo a drastically different evolution path from that of Enceladus. Due to its larger size and slower rotation rate, the ocean heat transport on Europa is likely to be much more efficient and thus the equilibrium ice thickness variation is predicted to be lower than 1~km, in line with the so-far available observations \citep{Iess-Stevenson-Parisi-et-al-2014:gravity, Beuthe-Rivoldini-Trinh-2016:enceladuss, Tajeddine-Soderlund-Thomas-et-al-2017:true, Cadek-Soucek-Behounkova-et-al-2019:long, Hemingway-Mittal-2019:enceladuss, Nimmo-Thomas-Pappalardo-et-al-2007:global}.

To drive the point home, the equilibrium ice geometry for Enceladus and Europa are solved by integrating an ice evolution model, where OHT is parameterized based on the scaling laws. All Europa scenarios with OHT parameterization form a rather flat ice shell with thickness variation below 2km. Most equilibrium ice geometries obtained using Enceladus parameters, to the contrary, exhibit strong thickness variations, unless the ice sheet is very mobile (low ice viscosity) or the assumed vertical diffusivity and the thermal expansion coefficient are both high. Some Enceladus scenarios even form the significant hemispheric asymmetry seen in observations \citep{Iess-Stevenson-Parisi-et-al-2014:gravity, Beuthe-Rivoldini-Trinh-2016:enceladuss, Tajeddine-Soderlund-Thomas-et-al-2017:true, Cadek-Soucek-Behounkova-et-al-2019:long, Hemingway-Mittal-2019:enceladuss}, indicating that the symmetry breaking mechanism proposed by \citet{Kang-Flierl-2020:spontaneous} can work in presence of OHT.

It should be noted that other factors, such as ice viscosity, vertical mixing in the ocean and thermal expansion coefficient (determined by ocean salinity), also have significant impacts on the equilibrium ice geometry for Enceladus. These factors are thus far poorly constrained. More work along this line will improve the prediction of equilibrium ice shell geometry.
Also, in this work, all heat is assumed to be produced in the ice shell. With heat produced in the silicate core, the ocean's stratification will change: a fresh ocean on a small moon with negative $\alpha$ will become more stratified, whereas a salty ocean on a large moon with positive $\alpha$ will become less stratified and even globally convective. In appearance of convection (salty/high pressure), the heating delivered to the ice shell by convective Taylor plumes is not going to be evenly distributed if the heating released from the seafloor is \citep{Soderlund-Schmidt-Wicht-et-al-2014:ocean, Soderlund-2019:ocean, Bire-Kang-Ramadhan-et-al-2022:exploring}. This may induce topography in the ice shell as well and needs to be investigated in the future. However, we expect the effect of bottom heating to play a less important role as the satellite size increases, because the vertical temperature gradient induced by bottom heating decrease with gravity \footnote{According to \citet{Gastine-Wicht-Aubert-2016:scaling}, Nusselt number is proportional to Rayleigh number to the power of $1.5$. Solving the vertical temperature gradient $\Delta T$ assuming fixed vertical heat flux yields $\Delta T \propto g^{-3/5}$.}, whereas the temperature difference induced by ice topography will increase with satellite size; and even for an icy moon as small as Enceladus, the vertical temperature gradient induced by a 40~mW/m$^2$ bottom heating is likely one order of magnitude smaller than that induced by the observed ice thickness variation \citep{Kang-Mittal-Bire-et-al-2021:how}. 

Despite these uncertainties, the qualitative result that ocean heat transport is more efficient at limiting ice-shell thickness variations on large satellites is likely to be robust. By connecting the equilibrium ice shell geometry with the icy moon's orbital parameters and the ocean properties, this work may bring a bit more constraints to the poorly-constrained icy worlds.

\begin{acknowledgments}
  This work is carried out in the Department of Earth, Atmospheric and Planetary Science (EAPS) in MIT. WK acknowledges support as a Lorenz-Houghton Fellow by endowed funds in EAPS and helpful comments from Prof. Malte Jansen.
\end{acknowledgments}

%



\software{MITgcm \citep{MITgcm-group-2010:mitgcm}}



\appendix
\section{A description of the General Circulation Model.}

  
  Our simulations are carried out using the Massachusetts Institute of Technology OGCM \citep[MITgcm][]{MITgcm-group-2010:mitgcm, Marshall-Adcroft-Hill-et-al-1997:finite} configured for application to icy moons. 
  
  The model integrates the non-hydrostatic primitive equations for an incompressible fluid in height coordinates, including a full treatment of the Coriolis force in a deep fluid, as described in \citet{MITgcm-group-2010:mitgcm, Marshall-Adcroft-Hill-et-al-1997:finite}. Such terms are typically neglected when simulating Earth's ocean because the ratio between the fluid depth and horizontal scale is small. Instead, when the moon size is order hundreds of kilometers like Enceladus, the aspect ratio is order $0.1$ and so not negligibly small. The size of each grid cell shrinks with depth due to spherical geometry and is accounted for by switching on the ``deepAtmosphere'' option of MITgcm. Also, the gravity will vary with depth as well. This is accounted for using the following profile of gravity.
  \begin{equation}
    \label{eq:g-z}
    g(z)=\frac{4\pi G\left[\rho_{\mathrm{core}}(a-D_0-H_0)^3+\rho_{\mathrm{out}}((a-z)^3-(a-D_0-H_0)^3)\right]}{3(a-z)^2}.
  \end{equation}
  In the above equation, $G=6.67\times10^{-11}$~N/m$^2$/kg$^2$ is the gravitational constant, $\rho_{\mathrm{core}}=2500$~kg/m$^3$ is the assumed core density and $\rho_{\mathrm{out}}=1000$~kg/m$^3$ is the density of the ocean/ice layer. $D_0$ and $H_0$ is the thickness of ocean and ice on global average.
  
  Since it takes several tens of thousands of years for our solutions to reach equilibrium, all of our experiments are first run under a zonally symmetric 2D configuration with a moderate resolution of $2$~degree ($8.7$~km). Only 30 layers (each 2~km) are used to keep the computational cost manageable. After equilibrium is reached, I interpolate the pick up files to generate initial conditions for the corresponding 3D simulation, which has a default horizontal resolution of 0.25$\times$0.25~degree and 70 unevenlly distributed vertical layers, whose thicknesses increase from 500~m to 2~km from top to bottom. Since changing rotation rate has significant impacts on the size of the jets and eddies, I had to adjust the grid width along east-west direction by a factor of 1.4 in those cases to better capture the dynamics. 

  \subsection{Diffusivity and Viscosity.}
  Vertical diffusivity affects the energetics of the ocean \citep{Young-2010:dynamic, Jansen-Kang-Kite-2022:energetics}. To account for the mixing of heat and salinity by unresolved turbulence, in our calculations, I set the explicit vertical diffusivity to $0.001$~m$^2$/s in both 2D and 3D simulations, following \citet{Kang-Mittal-Bire-et-al-2021:how}. This is roughly 4 orders of magnitude greater than molecular diffusivity, but broadly consistent with dissipation rates suggested by \citet{Rekier-Trinh-Triana-et-al-2019:internal} for Enceladus, according to the scaling suggested by \citet{Wunsch-Ferrari-2004:vertical}. In all experiments, horizontal diffusivity is set to be equal to the vertical diffusivity regardless of the resolution and the size of the icy moon.

  Viscosity is necessary to keep the model stable. For the coarse resolution 2D simulations, the horizontal viscosity are set to $\frac{a}{150~\mathrm{km}}$~m$^2$/s ($a$ is the radius of the moon). Additionally, a bi-harmonic hyperviscosity of $\frac{a}{150~\mathrm{km}}\times 10^8$~m$^4$/s is employed to further damp numerical noise induced by our use of stair-like ice topography. The same viscosities are used in \citet{Kang-Jansen-2022:in}. For the high resolution 3D simulations, the explicit horizontal and vertical viscosity is set to much smaller values (0.08~m$^2$/s and 0.03~m$^2$/s respectively), and I use the widely-applied Smagorinsky viscosity scheme \citep{Smagorinsky-1963:general}. Unlike the fixed viscosity scheme, Smagorinsky scheme determines the viscosity based on the resolved dynamics, and as a result, numerical noise will be damped while dynamics can be kept to a larger extent. The Smagorinsky viscosity constant is set to $3$ by default. As mentioned before, resolution in the x-direction is increased (decreased) by a factor of 1.4 under a higher (lower) rotation rate. In those experiments, horizontal viscosity is adjusted proportional to the x-grid width.
  
  In the coarse resolution model, convection cannot be resolved, so parameterization is needed. Following \citet{Kang-Mittal-Bire-et-al-2021:how}, I set the diffusivity to a much larger value in convectively unstable regions, to represent the vertical mixing associated with convective overturns. This convective diffusivity $\kappa_{\mathrm{conv}}$ is set to increase from $1$~m$^2$/s to $30$~m$^2$/s as gravity and convection strengthens with the satellite radius. Similar approaches are widely used to parameterize convection in coarse resolution ocean models (see, e.g. \citet{Klinger-Marshall-Send-1996:representation}) and belong to a family of convective adjustment schemes. Our results turn out to be insensitive to $\kappa_{\mathrm{conv}}$, as long as the convective timescale $D^2/\kappa_{\mathrm{conv}}< 1$~yr is much shorter than the advective time scale $M_{\mathrm{half}}/\Psi\approx 10^2$-$10^3$~yrs ($M_{\mathrm{half}}$ is half of the total mass of the ocean and $\Psi$ is the maximum meridional streamfunction in $kg/s$).

  \subsection{Equation of state and the freezing point of water}
  To make the dynamics as realistic as possible, the ``MDJWF'' equation of state \citep[EOS][]{McDougall-Jackett-Wright-et-al-2003:accurate} is adopted when it comes to determine density using temperature, salinity and pressure. As demonstrated by Fig.~1e of the main text, the thermal expansion coefficient $\alpha$ at the freezing point is negative at the ice-ocean interface when the moon size is small (low pressure) and when the ocean is fresh. This anomalous expansion can suppress the convection driven by bottom heating \citep{Zeng-Jansen-2021:ocean, Kang-Marshall-Mittal-et-al-2022:dynamics} and can alter the direction of ocean circulation \citep{Kang-Mittal-Bire-et-al-2021:how}.
  
  The freezing point of water $T_f$ is assumed to depend on local pressure $P$ and salinity $S$ as follows,
  \begin{equation}
    \label{eq:freezing-point}
    T_f(S,P)=c_0+b_0P+a_0S,
  \end{equation}
where $a_0=-0.0575$~K/psu, $b_0=-7.61\times10^{-4}$~K/dbar and $c_0=0.0901$~degC. The pressure $P$ can be calculated using hydrostatic balance $P=\rho_igH$ ($\rho_i=917$~kg/m$^3$ is the density of the ice and $H$ is the ice thickness).

  \subsection{Boundary conditions}
  \label{sec:boundary-conditions}

  The ocean is encased by an ice shell with meridionally-varying thickness, assuming hydrostacy (i.e., ice is floating freely on the water). The ice thickness is set to be
  \begin{equation}
    \label{eq:Hice}
    H(\phi)=H_0-H_2P_2(\sin\phi),
  \end{equation}
  where $H_0$ is the mean ice thickness, $P_2$ is the 2nd order Legendre polynomial, and $H_2$ is the amplitude of the ice thickness variation. $\phi$ denotes latitude. The thickness profile is shown by a solid curve in Fig.1b of the main text. Partial cells is switched on to better represent the ice topography: water is allowed to occupy a fraction of the height of a whole cell with an increment of 10\%. Interactions between the ice shell and the ocean is taken care of by a modified version of the MITgcm's ``shelfice'' module \citep{Losch-2008:modeling}. The ocean is forced by heat and salinity fluxes from the ice shell at the top.
  
  \underline{Diffusion of heat through the ice}
  
  Heat loss to space by heat conduction through the ice $\mathcal{H}_{\mathrm{cond}}$ is represented using a 1D vertical heat conduction model,
\begin{equation}
  \mathcal{H}_{\mathrm{cond}}=\frac{\kappa_{0}}{H} \ln \left(\frac{T_{f}}{T_{s}}\right),
  \label{eq:H-cond}
  \end{equation}
  where $H$ is the thickness of ice (solid curve in Fig.1b of the main text), the surface temperature is $T_s$ and the ice temperature at the water-ice interface is the local freezing point $T_f$ (Eq.~\ref{eq:freezing-point}).
  The surface temperature $T_s$ is set to the radiative equilibrium temperature, which can be computed given the incoming solar radiation and obliquity ($\delta=3^\circ$) and assuming an albedo of $0.81$. Typical heat losses averaged over the globe are $\mathcal{H}_{\mathrm{cond}}$= $50$~mW/m$^2$, broadly consistent with observations \citep{Tajeddine-Soderlund-Thomas-et-al-2017:true}.

  \underline{Ice-ocean fluxes}

  The interaction between ocean and ice is simulated using MITgcm's ``shelf-ice'' package \citep{Losch-2008:modeling, Holland-Jenkins-1999:modeling} with some modifications. 

  At the water-ice interface, we consider the response of the ocean to a prescribed ice freezing rate while ignoring the possible response of the ice to the water-ice heat/salinity exchange. The freezing/melting induces a salinity/fresh water flux into the ocean (we assume the ice salinity to be zero); meanwhile, the ocean temperature at the upper boundary is relaxed to the local freezing point $T_f$ determined by the local salinity and pressure (Eq.~\ref{eq:freezing-point}).
  \begin{eqnarray} \frac{dS_{\mathrm{ocn-top}}}{dt}&=&\frac{qS_{\mathrm{ocn-top}}}{\delta z}\label{eq:S-tendency}\\
    \frac{dT_{\mathrm{ocn-top}}}{dt}&=&\frac{1}{\delta z}(\gamma_T-q)(T_f-T_{\mathrm{ocn-top}})\label{eq:T-tendency}
  \end{eqnarray}
  Here, $S_{\mathrm{ocn-top}}$ and $T_{\mathrm{ocn-top}}$ denote the upper boundary salinity and temperature, $\gamma_T=10^{-5}$~m/s are the water-ice exchange coefficients for temperature and salinity, $\delta z=2$~km is the thickness of the water-ice ``boundary layer'' and $q$ is the freezing rate in m/s (note that $q$ is orders of magnitude smaller than $\gamma_T$). The ``boundary layer'' option is switched on to avoid possible numerical instabilities induced by an ocean layer which is too thin. 
     
  In addition to the above conditions on temperature and salinity, the tangential velocity is relaxed back to zero at a rate of $\gamma_M=10^{-3}$m/s at the upper and lower boundaries.

 \subsection{Ice flow model}

  The prescribed freezing rate $q$ is computed using the divergence of the ice flow, assuming the ice sheet geometry is in equilibrium. Here, an upside-down land ice sheet model is used following \citet{Ashkenazy-Sayag-Tziperman-2018:dynamics}. The ice flows down its thickness gradient, driven by the pressure gradient induced by the spatial variation of the ice top surface, somewhat like a second order diffusive process. At the top, the speed of the ice flow is negligible because the upper part of the shell is so cold and hence rigid; at the bottom, the vertical shear of the ice flow speed vanishes, as required by the assumption of zero tangential stress there. This is the opposite to that assumed in the land ice sheet model. In rough outline, I calculate the ice flow using the expression below obtained through repeated vertical integration of the force balance equation (the primary balance is between the vertical flow shear and the pressure gradient force), using the aforementioned boundary conditions to arrive at the following formula for ice transport $\mathcal{Q}$,
\begin{equation}
  \mathcal{Q}(\phi)= \mathcal{Q}_0H^3(\partial_\phi H/a) \label{eq:ice-flow}
\end{equation}
where
\begin{equation}
\mathcal{Q}_0=\frac{2(\rho_0-\rho_i)g}{\eta_{m}(\rho_0/\rho_i)\log^3\left(T_f/T_s\right)}\int_{T_s}^{T_f}\int_{T_s}^{T(z)}\exp\left[-\frac{E_{a}}{R_{g} T_{f}}\left(\frac{T_{f}}{T'}-1\right)\right]\log(T')~\frac{dT'}{T'}~\frac{dT}{T}.\nonumber 
\end{equation}
Here, $\phi$ denotes latitude, $a$ and $g$ are the radius and surface gravity of the moon, $T_s$ and $T_f$ are the temperature at the ice surface and the water-ice interface (equal to local freezing point, Eq.~\ref{eq:freezing-point}), and $\rho_i=917$~kg/m$^3$ and $\rho_0$ are the ice density and the reference water density. $E_a=59.4$~kJ/mol is the activation energy for diffusion creep, $R_g=8.31$~J/K/mol is the gas constant and $\eta_{m}$ is the ice viscosity at the freezing point. The latter has considerable uncertainty \citep[$10^{13}$-$10^{16}$~Pa$\cdot$s][]{Tobie-Choblet-Sotin-2003:tidally}, and here $\eta_{m}$ is set to $10^{14}$~Pa$\cdot$s.

In steady state, the freezing rate $q$ must equal the divergence of the ice transport thus:
\begin{equation}
    q=-\frac{1}{a\cos\phi}\frac{\partial}{\partial \phi} (Q\cos\phi).
    \label{eq:freezing-rate}
\end{equation}
As shown by the dashed curve in Fig.1b of the main text, ice melts in high latitudes and forms in low latitudes at a rate of a few kilometers every million years. A more detailed description of the ice flow model can be found in \citet{Kang-Flierl-2020:spontaneous} and \citet{Ashkenazy-Sayag-Tziperman-2018:dynamics}. Freezing and melting leads to changes in local salinity and thereby a buoyancy flux. 

\subsection{Model of tidal dissipation in the ice shell}
\label{sec:tidal-dissipation-model}

Icy moon's ice shell is periodically deformed by tidal forcing and the resulting strains in the ice sheet produce heat. I follow \citet{Beuthe-2019:enceladuss} to calculate the ice dissipation rate. Instead of repeating the whole derivation here, I only briefly summarize the procedure and present the final result. Unless otherwise stated, parameters are the same as assumed in \citet{Kang-Flierl-2020:spontaneous}.

Tidal dissipation consists of three components \citep{Beuthe-2019:enceladuss}: a membrane mode $\mathcal{H}_{\mathrm{ice}}^{\mathrm{mem}}$ due to the extension/compression and tangential shearing of the ice membrane, a mixed mode $\mathcal{H}_{\mathrm{ice}}^{mix}$ due to vertical shifting, and a bending mode $\mathcal{H}_{\mathrm{ice}}^{bend}$ induced by the vertical variation of compression/stretching. Following \citet{Beuthe-2019:enceladuss}, I first assume the ice sheet to be completely flat. By solving the force balance equation, I obtain the auxiliary stress function $F$, which represents the horizontal displacements, and the vertical displacement $w$. The dissipation rate $\mathcal{H}_{\mathrm{ice}}^{\mathrm{flat,x}}$ (where $x=\{\mathrm{mem},\mathrm{mix},\mathrm{bend}\}$ ) can then be written as a quadratic form of $F$ and $w$. In the calculation, the ice properties are derived assuming a globally-uniform surface temperature of 60K and a melting viscosity of $5\times10^{13}$~Pa$\cdot$s. 

Ice thickness variations are accounted for by multiplying the membrane mode dissipation $\mathcal{H}_{\mathrm{ice}}^{\mathrm{flat,mem}}$, by a factor that depends on ice thickness. The membrane mode is the only mode which is amplified in thin ice regions (see \citet{Beuthe-2019:enceladuss}). This results in the expression:
\begin{equation}
  \label{eq:H-tide}
  \mathcal{H}_{\mathrm{ice}}=(H/H_0)^{p_\alpha}\mathcal{H}_{\mathrm{ice}}^{\mathrm{flat,mem}}+\mathcal{H}_{\mathrm{ice}}^{\mathrm{flat,mix}}+\mathcal{H}_{\mathrm{ice}}^{\mathrm{flat,bend}},
\end{equation}
where $H$ is the prescribed thickness of the ice shell as a function of latitude and $H_0$ is the global mean of $H$. Since thin ice regions deform more easily and produce more heat, $p_\alpha$ is negative. Because more heat is produced in the ice shell, the overall ice temperature rises, which, in turn, further increases the mobility of the ice and leads to more heat production (the rheology feedback).


The tidal heating profile corresponding to $p_\alpha=-1.5$ is the red solid curve plotted in Fig.1c of the main text.

\section{Idealized Ice evolution model.}
Here I provide a brief overview for the idealized model used to evolve the ice shell of Enceladus and Europa. Interested readers are referred to \citep{Kang-Flierl-2020:spontaneous} and its supplementary material for more detail.

In this model, ice shell thickness $H$ is changed over time by the melting induced by the tidal heating $\mathcal{H}_{\mathrm{ice}}$ (given by Eq.~\ref{eq:H-tide}), the down-gradient ice flow $\mathcal{Q}$ (given by Eq.~\ref{eq:ice-flow}), the heat loss to space by conduction $\mathcal{H}_{\mathrm{cond}}$ (given by Eq.~\ref{eq:H-cond}), the crack-induced cooling $\mathcal{H}_{\mathrm{crack}}$ in places where ice is sufficiently thin, and heat transmitted upward by the ocean $\mathcal{H}_{\mathrm{ocn}}$. The ice thickness tendency can be symbolically expressed as following,
\begin{equation}
  \label{eq:ice-evolution}
  \frac{dH}{dt}=\frac{\mathcal{H}_{\mathrm{cond}}(H)-\mathcal{H}_{\mathrm{ice}}(H)-\mathcal{H}_{\mathrm{ocn}}}{L_f\rho_i } + \frac{1}{a\sin\phi}\partial_\phi \left(\sin\phi \mathcal{Q}(H)\right),
\end{equation}
where $L_f$ and $\rho_i$ are the latent heat of freezing and density for ice, $a$ is the moon's radius and $\phi$ denotes latitude.
 Physical constants and parameters for Enceladus and Europa are stated in Table.\ref{tab:parameters-two-moons}.
$\mathcal{H}_{\mathrm{ice}}$ is polar-amplified, and as a result, the polar ice shell tends to be thinner, which in turn increases the heat production over the pole (see Eq.~\ref{eq:H-tide}). The tendency for ice thickness variation to increase due to the rheology feedback will be balanced by the rapid heat loss through thin ice (Eq.~\ref{eq:H-cond}) and the transport by ice flow (Eq.~\ref{eq:ice-flow}). An additional heat sink is activated only when the ice thickness is less than $H_{\mathrm{crack}}=3$~km to prevent further melting, and crudely represents the effect of cracks and geysers that carry the extra heat away. For all time, the global tidal dissipation $\mathcal{H}_{\mathrm{ice}}$ is scaled to exactly balance the instantaneous conductive heat loss $\mathcal{H}_{\mathrm{cond}}$. By so doing, the rheology feedback and thus the ice thickness variation are maximized. Throughout the integration, the global mean ice thickness is fixed at $H_0=20$~km.

The ocean-ice heat exchange is a new process I introduced. Inspired by the conceptual model, the heat flux coming from the ocean is parameterized by Eq.~19 in the main text. 

The initial condition is set as follows
\begin{equation}
  \label{eq:ice-evolution-IC}
  H(0)=H_0-H_2P_2(\sin\phi)-H_1P_1(\sin\phi),
\end{equation}
where $H_0=20$~km, $H_2=3$~km and $H_1=1$~km. $P_1$ and $P_2$ are the first and second order of Legendre Polynomials. 


\begin{table*}[hptb!]
  
  \centering
  \begin{tabular}{lll}
    Symbol & Name & Definition/Value\\
    \hline
    \multicolumn{3}{c}{Physical constants}\\
    \hline
    $L_f$ & fusion energy of ice & 334000~J/kg\\
    $C_p$ & heat capacity of water & 4000~J/kg/K\\
    $T_f(S,P)$ & freezing point & Eq.\ref{eq:freezing-point}\\
    $\rho_i$ & density of ice & 917~kg/m$^3$ \\
    $\rho_w$ & density of the ocean & 'MDJWF' scheme \cite{McDougall-Jackett-Wright-et-al-2003:accurate} \\
    $\alpha$ & thermal expansion coeff. &  $-\partial (\rho/\rho_0)/\partial T$ \\
    $\beta$ & saline contraction coeff. &  $\partial (\rho/\rho_0)/\partial S$\\
    $\kappa_0$ & conductivity coeff. of ice & 651~W/m\\
    $p_\alpha$ & ice dissipation amplification factor & -1.5 \\
    $\eta_{m}$ & ice viscosity at freezing point & 10$^{14}$~Ps$\cdot$s\\
    \hline
    \multicolumn{3}{c}{Default model setup}\\
    \hline
    $a$ & radius & 2500~km\\
    $g_0$ & surface gravity & Eq.~\eqref{eq:g-z}\\
    $\delta$ & obliquity & 3.1$^\circ$\\
    $H_0$ & global mean ice thickness & 20~km  \\
    $H_2$ & equator-to-pole ice thickness variation & 3~km\\
    $H$ & ice shell thickness & Eq.\ref{eq:Hice}\\
    $D$ & global mean ocean depth& 56~km \\
    $\Omega$ & rotation rate & 2.05$\times$10$^{-5}$~s$^{-1}$\\
    $\bar{T_s}$ & mean surface temperature& 110K\\
    $S_0$ & mean ocean salinity & 60~psu\\
    $P_0$ & reference pressure & $\rho_ig_0H_0$ \\
    $T_0$ & reference temperature & $T_f(S_0,P_0)$ \\
    $\nu_h$ & horizontal viscosity & 0.08~m$^2$/s (3D), $(a/150$~km$)$~m$^2$/s (2D)\\
    $\nu_v$ & vertical viscosity & 0.03~m$^2$/s (3D), 1~m$^2$/s (3D)\\
    $\nu_{\mathrm{smag}}$ & Smagorinsky viscosity (3D only) & 3 \\
    $\tilde{\nu}_h,\ \tilde{\nu}_v$ & bi-harmonic hyperviscosity (2D only) & $\frac{a}{150~\mathrm{km}}\times$10$^8$~m$^4$/s\\ 
    $\kappa_h,\ \kappa_v$ & horizontal/vertical diffusivity & 0.001~m$^2$/s\\
    $(\gamma_T,\ \gamma_S,\ \gamma_M)$ & water-ice exchange coeff. for T, S \& momentum & (10$^{-5}$, 10$^{-5}$, 10$^{-4}$)~m/s\\
    $\mathcal{H}_{\mathrm{cond}}$ & conductive heat loss through ice & Eq.\ref{eq:H-cond}\\
    $\mathcal{H}_{\mathrm{ice}}$ & tidal heating produced in the ice & Eq.\ref{eq:H-tide} \\
    \hline
     \end{tabular}
  \caption{Model parameters used in the ocean general circulation model and the conceptual model. }
  \label{tab:parameters}
  
\end{table*}

\begin{table*}[hptb!]
  
  \centering
  \begin{tabular}{lll}
    Symbol & Name & Definition/Value\\
    \hline
    $p_\alpha$ & ice dissipation amplification factor & -1.5\\
\hline
    \multicolumn{3}{c}{Parameters for Enceladus}\\
    \hline
    $a$ & radius & 252~km\\
    $g_0$ & surface gravity & 0.113~m/s$^2$\\
    $\delta$ & obliquity & 27$^\circ$\\
    $H_0$ & global mean ice thickness &  20~km \cite{Hemingway-Mittal-2019:enceladuss} \\
    $H_2$ & initial equator-to-pole ice thickness variation & 3~km\\
    $H_1$ & initial hemispherical asymmetry & 1~km\\
    $\bar{T_s}$ & mean surface temperature & 59K \\
    \hline
    \multicolumn{3}{c}{Parameters for Europa}\\
    \hline
    $a$ & radius & 1561~km\\
    $g_0$ & surface gravity & 1.315~m/s$^2$\\
    $\delta$ & obliquity & $3.1^\circ$\\
    $H_0$ & global mean ice thickness &  20~km \\
    $H_2$ & initial equator-to-pole ice thickness variation & 3~km\\
    $H_1$ & initial hemispherical asymmetry & -1~km\\
    $\bar{T_s}$ & mean surface temperature & 110K \\
    \hline
     \end{tabular}
  \caption{Parameters for Enceladus and Europa. }
  \label{tab:parameters-two-moons}
  
\end{table*}

 \begin{figure*}[htbp!]
    \centering
    \includegraphics[width=0.8\textwidth]{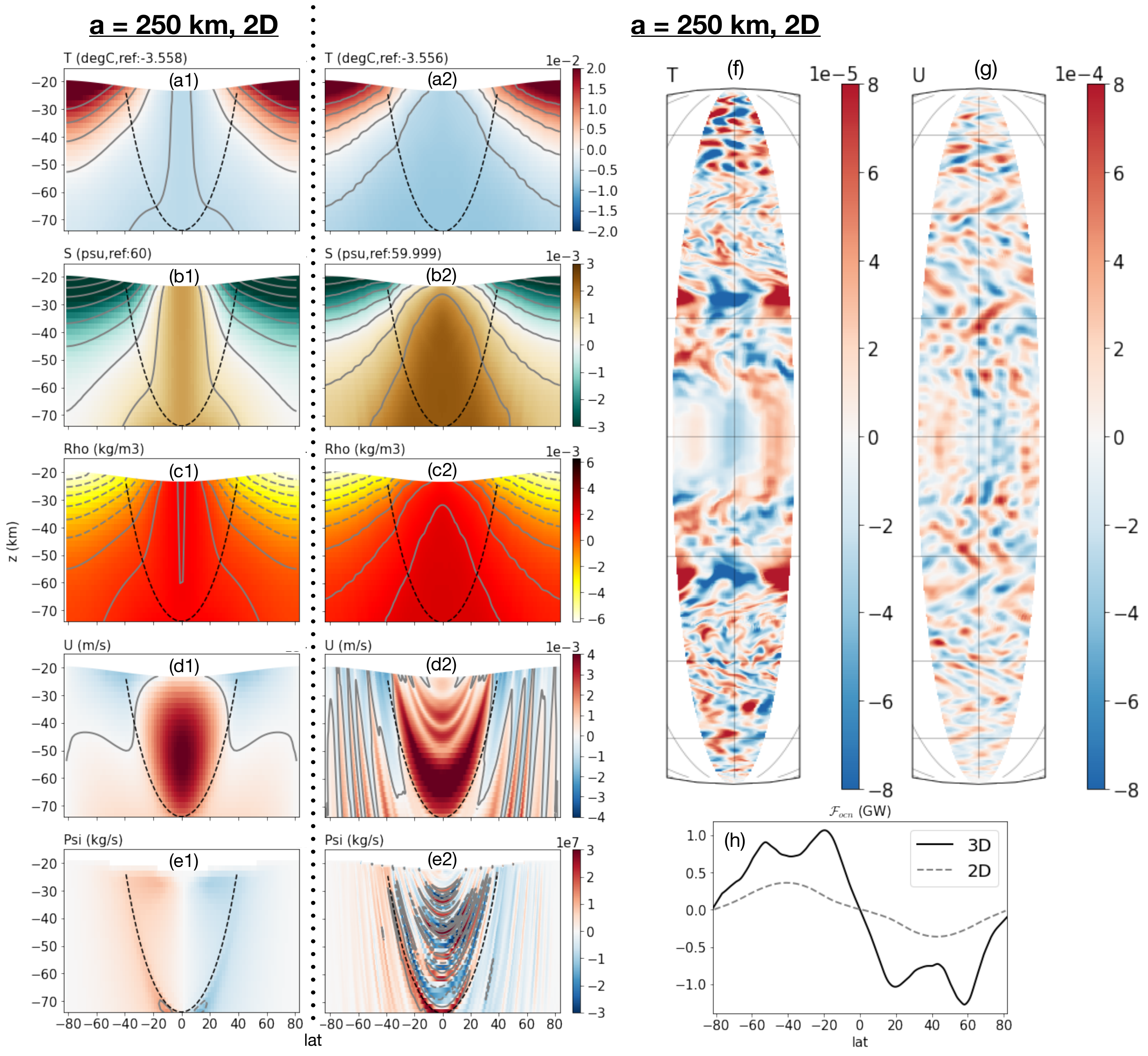}
    \caption{\small{Same as Fig.~\ref{fig:3d-dynamics} except for $a=250$~km instead of $2500$~km. }}
    \label{fig:3d-dynamics-a250}
  \end{figure*}

   \begin{figure*}[htbp!]
    \centering
    \includegraphics[width=0.8\textwidth]{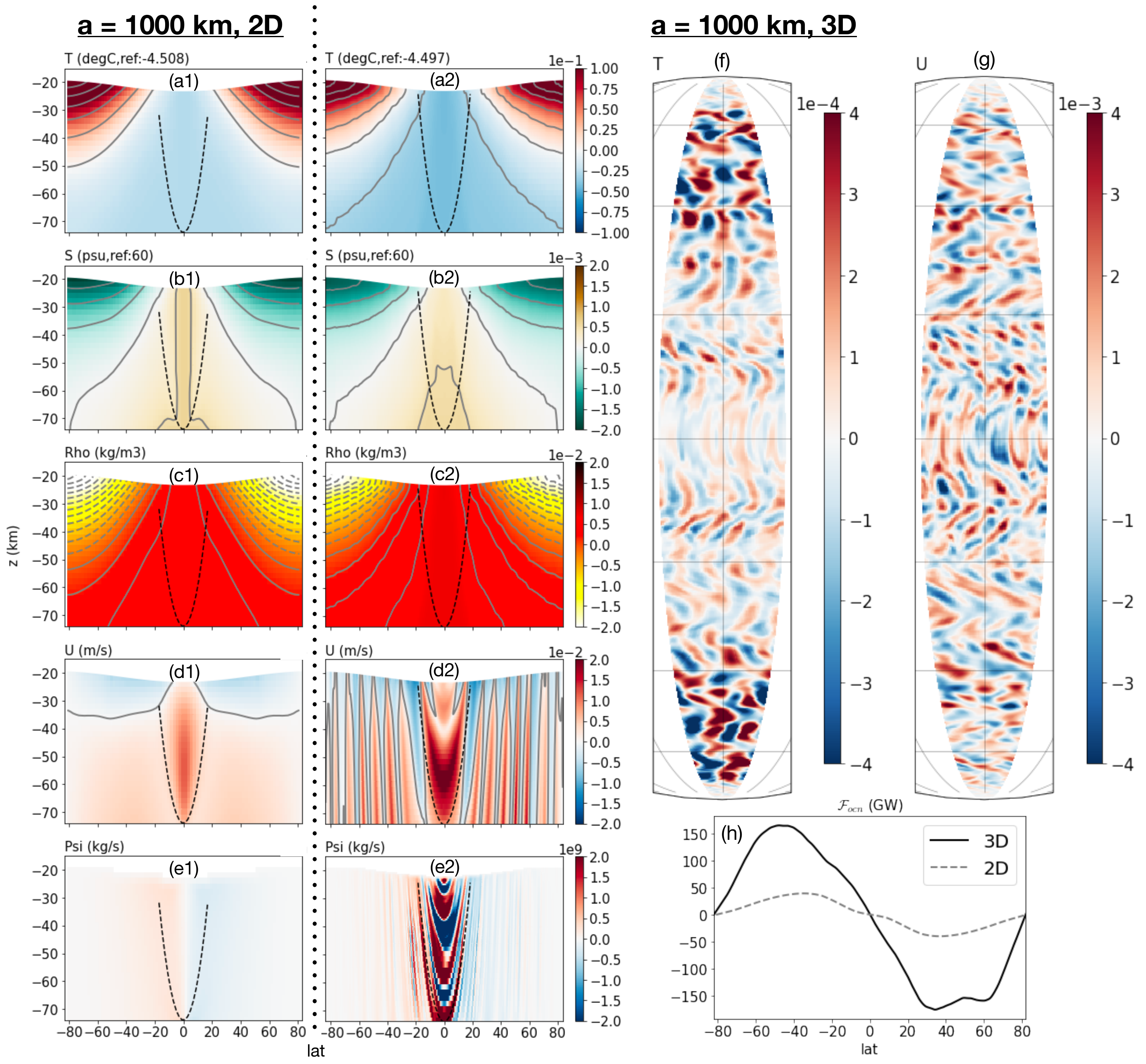}
    \caption{\small{Same as Fig.~\ref{fig:3d-dynamics} except for $a=1000$~km instead of $2500$~km. }}
    \label{fig:3d-dynamics-a1000}
  \end{figure*}

  \begin{figure*}[htbp!]
    \centering
    \includegraphics[width=0.8\textwidth]{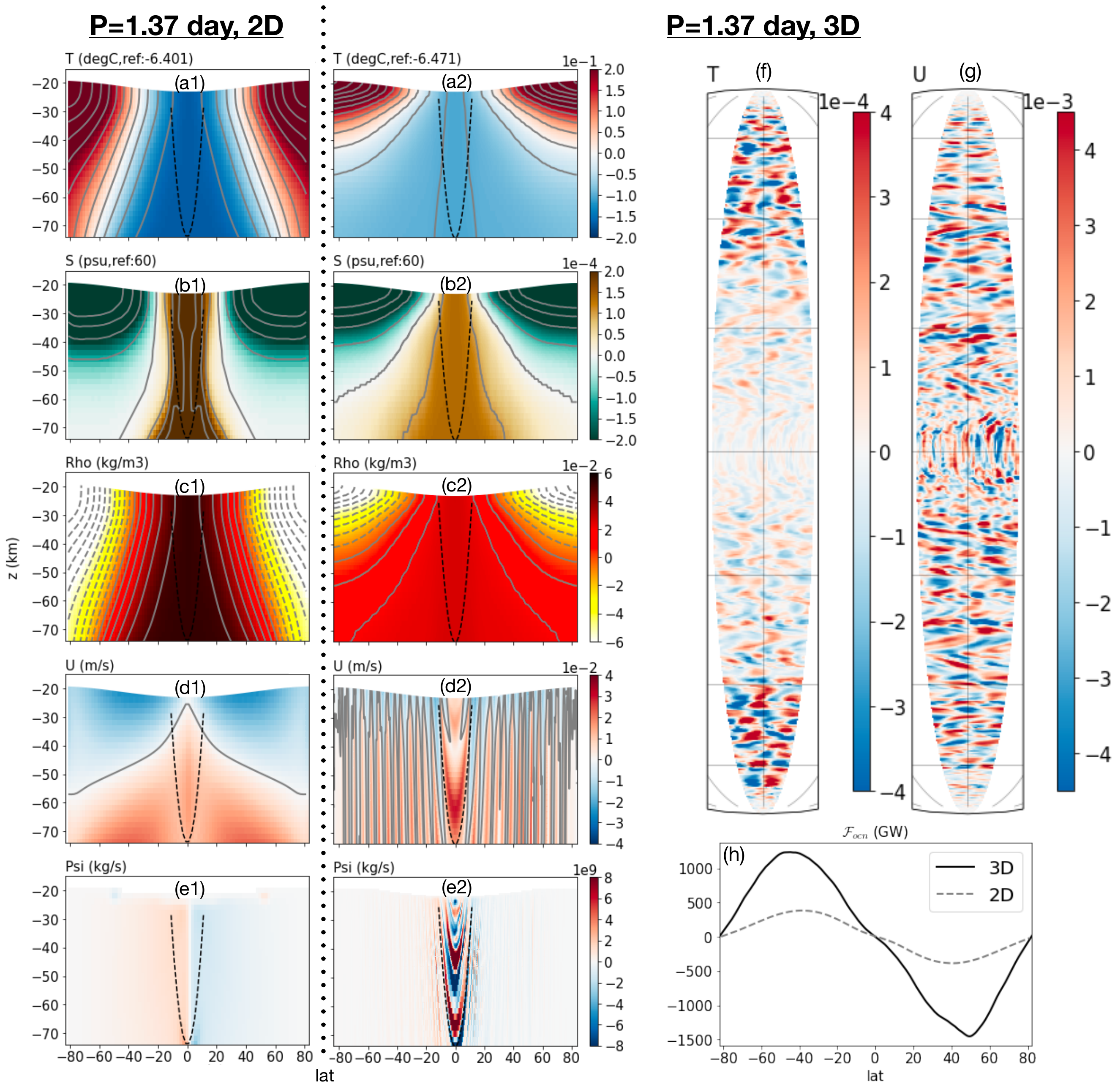}
    \caption{\small{Same as Fig.~\ref{fig:3d-dynamics} except rotation period is set to 1.37~day instead of 3.5~day. }}
    \label{fig:3d-dynamics-O1.37}
  \end{figure*}

    \begin{figure*}[htbp!]
    \centering
    \includegraphics[width=0.8\textwidth]{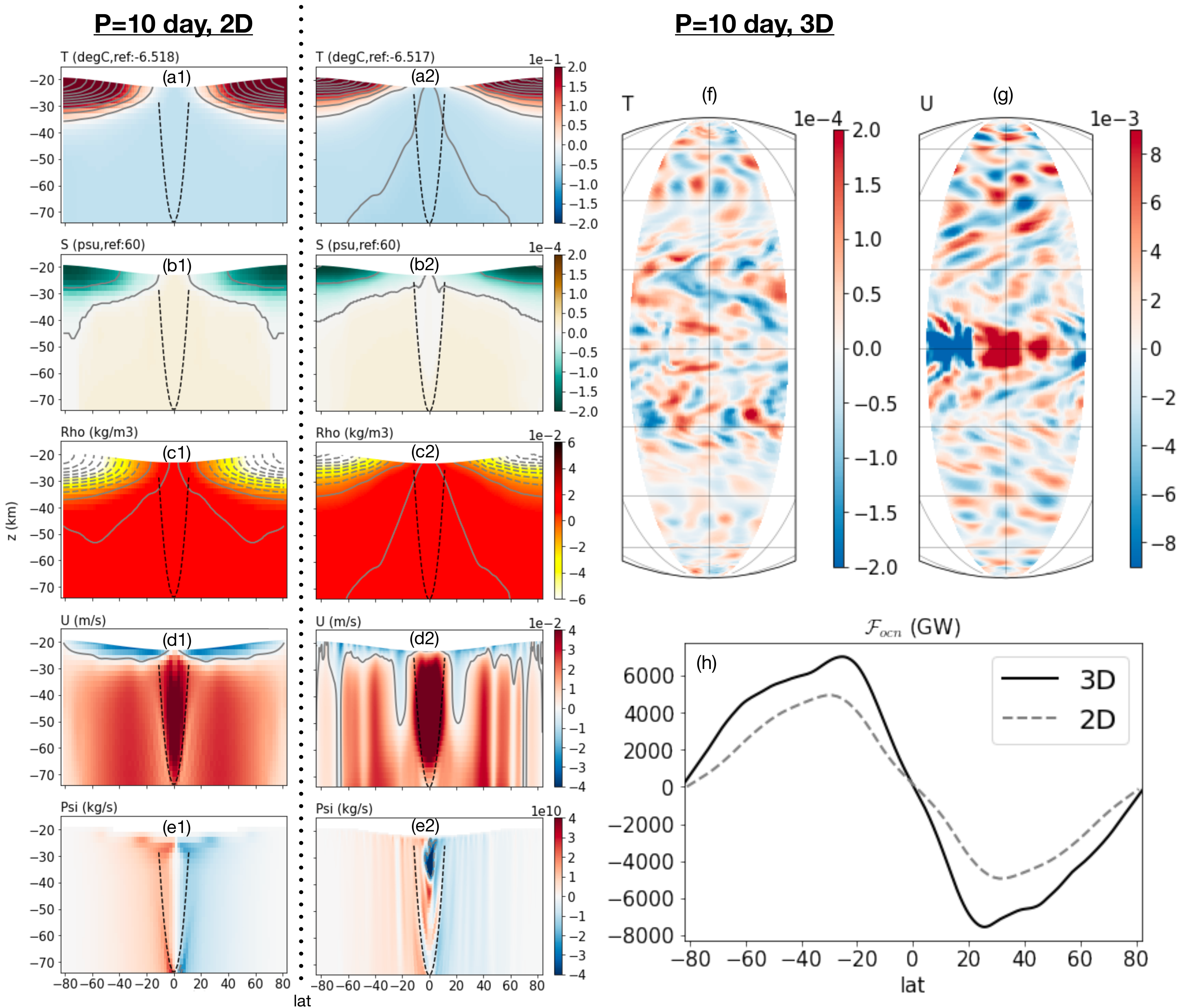}
    \caption{\small{Same as Fig.~\ref{fig:3d-dynamics} except rotation period is set to 10~day instead of 3.5~day. }}
    \label{fig:3d-dynamics-O10}
  \end{figure*}

 \begin{figure*}[htbp!]
    \centering
    \includegraphics[width=0.8\textwidth]{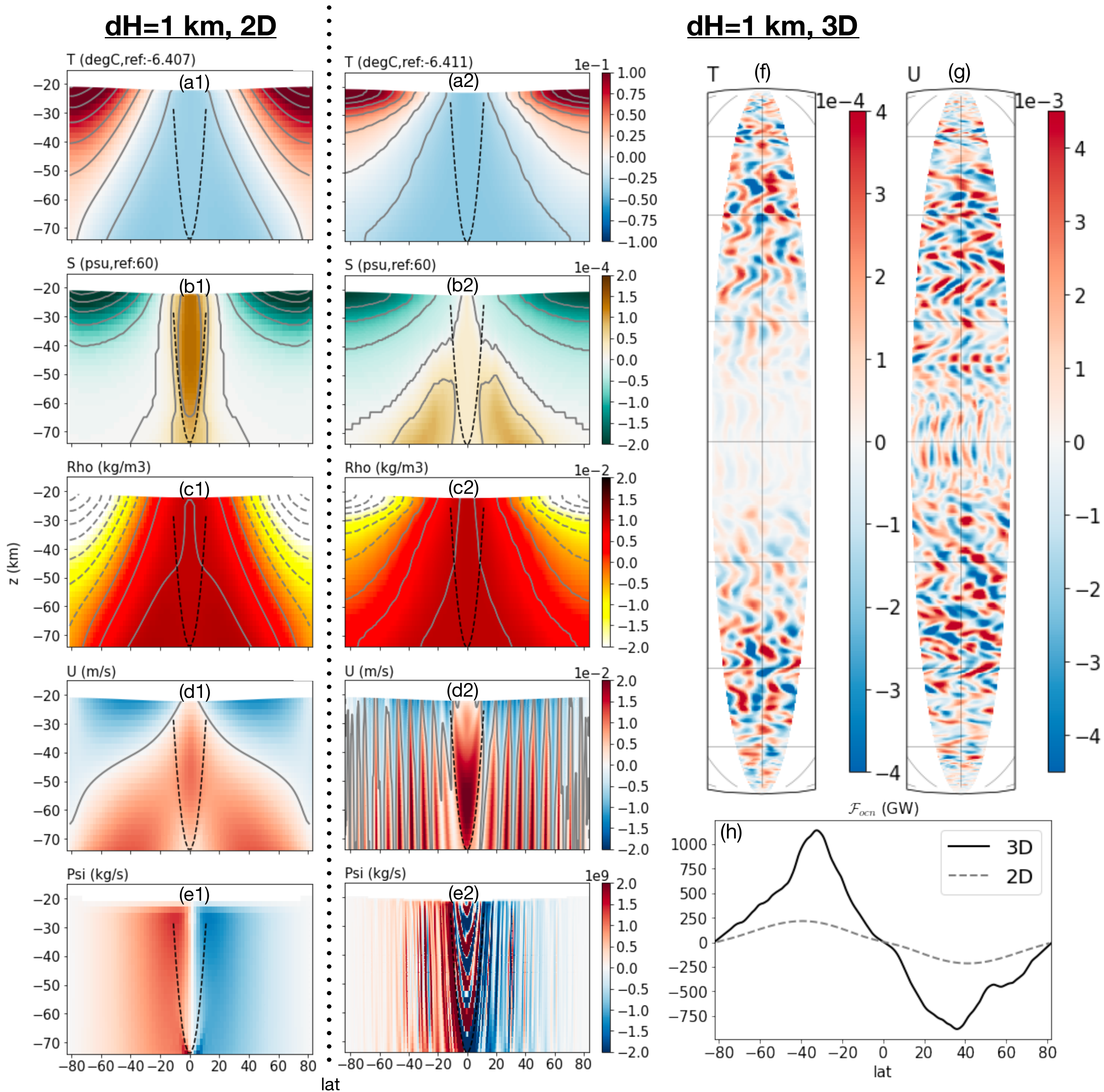}
    \caption{\small{Same as Fig.~\ref{fig:3d-dynamics} except the ice thickness contrast $\Delta H=1$~km instead of $3$~km. }}
    \label{fig:3d-dynamics-H1}
  \end{figure*} 

 \begin{figure*}[htbp!]
    \centering
    \includegraphics[width=0.8\textwidth]{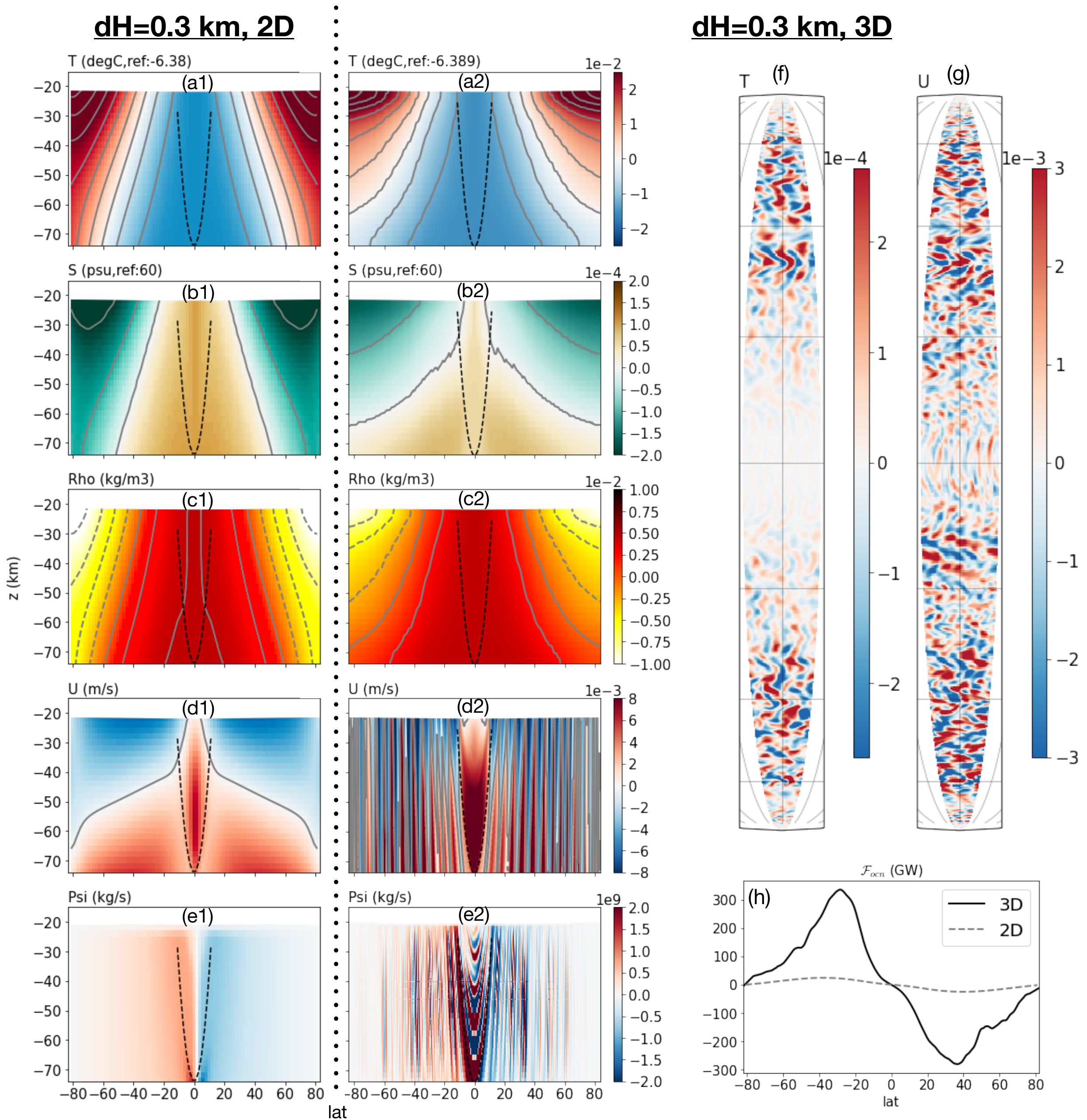}
    \caption{\small{Same as Fig.~\ref{fig:3d-dynamics} except the ice thickness contrast $\Delta H=0.3$~km instead of $3$~km. }}
    \label{fig:3d-dynamics-H03}
  \end{figure*}

    \begin{figure*}[htbp!]
    \centering
    \includegraphics[width=0.8\textwidth]{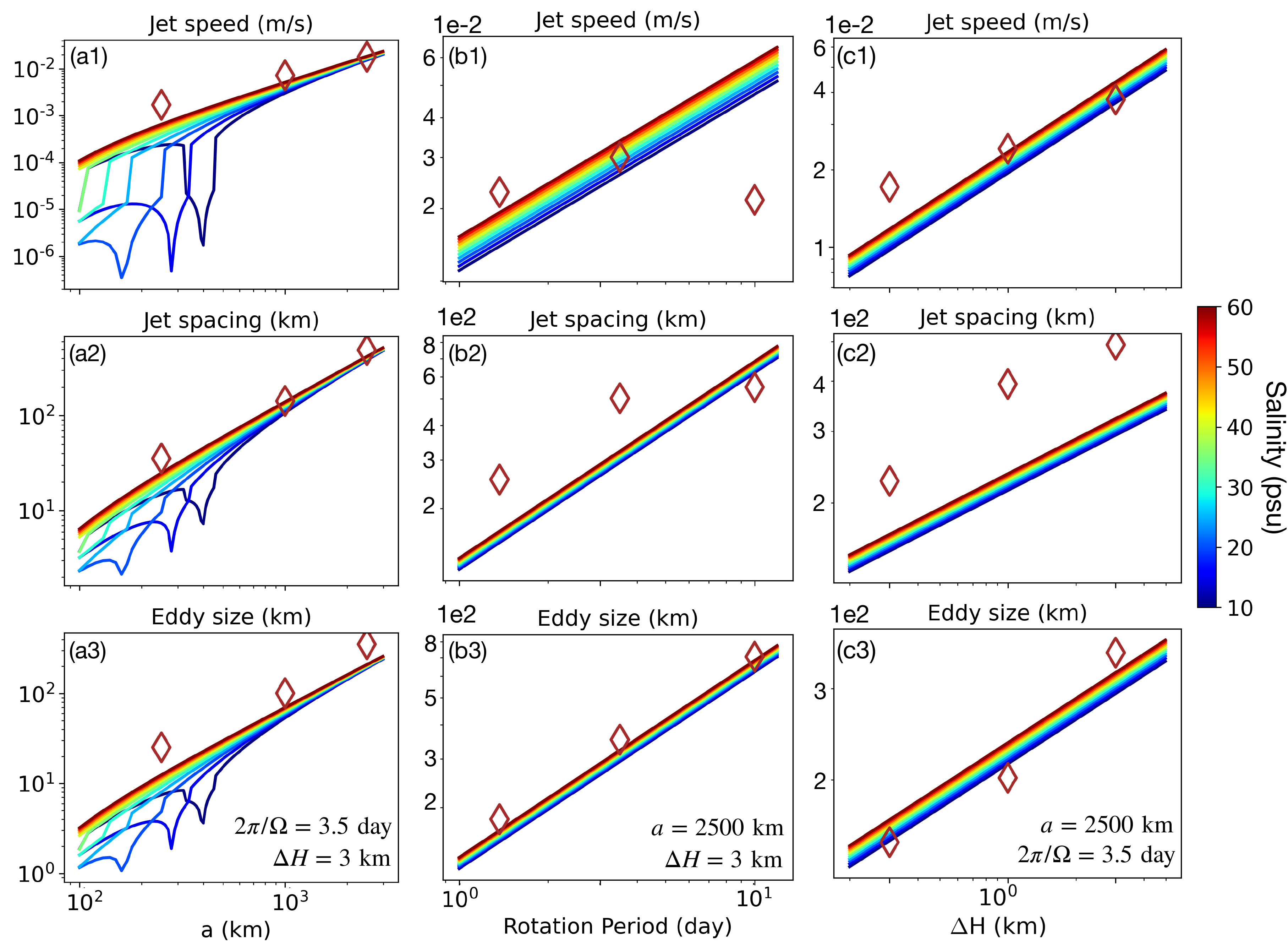}
    \caption{\small{Similar to Fig.~\ref{fig:eddy-scaling} except for scalings of eddy and jet properties. Top row shows jet speed, middle row shows jet spacing, and bottom row shows eddy size. Solid lines in the top row show the predicted jet speed given by Eq.~\eqref{eq:jet-xi-eq-1} and Eq.~\eqref{eq:U-xi-gt-1}, and solid lines in the bottom and lower rows show the Rhine scale given by Eq.~\eqref{eq:deformation-rhines-xi-eq-1} and Eq.~\eqref{eq:rhine-xi-gt-1} multiplied by a factor of $4\pi$. From blueish color to reddish color denotes increasing ocean salinity. }}
    \label{fig:eddy-scaling-2}
  \end{figure*}

\bibliography{export}{}
\bibliographystyle{aasjournal}



\end{document}